\documentclass[aps,prb,twocolumn]{revtex4-1}

\usepackage[english]{babel}
\usepackage[utf8]{inputenc}
\usepackage{amsmath}
\usepackage{amssymb}
\usepackage[caption=false]{subfig}
\usepackage{amssymb}
\usepackage{epsfig}
\usepackage{graphicx}
\usepackage{amsmath}
\usepackage{array,color}
\usepackage{natbib}

\usepackage{soul}
\usepackage{textcomp}
\usepackage[T1]{fontenc}

\usepackage[usenames,dvipsnames]{xcolor}
\definecolor{forestgreen}{rgb}{0.11,0.54,0.15}
\definecolor{purple}{rgb}{0.62,0.10,0.96}
\definecolor{dockerblue}{rgb}{0.11,0.56,0.98}
\definecolor{freeblue}{rgb}{0.25,0.41,0.88}

\usepackage[pdftex,plainpages=false,colorlinks=true,linkcolor=Red, citecolor=blue, urlcolor=blue]{hyperref}

\usepackage{latexsym}
\usepackage{url}
\usepackage{xcolor}

\usepackage{bm}

\usepackage{etoolbox}

\usepackage{appendix}

\begin{document}

\title{Resolving orientation-specific diffusion-relaxation features via Monte-Carlo density-peak clustering in heterogeneous brain tissue}

\author{A. Reymbaut$^{1,2}$}
\email{alexis.reymbaut@fkem1.lu.se}
\author{J. P. de Almeida Martins$^{1,2}$}
\author{C. M. W. Tax$^{3}$}
\author{F. Szczepankiewicz$^{4,5}$}
\author{D. K. Jones$^{6}$}
\author{D. Topgaard$^{1,2}$}

\affiliation{
$^1$Department of Physical Chemistry, Lund University, Lund, Sweden\\
$^2$Random Walk Imaging AB, Lund, Sweden\\
$^3$Cardiff University Brain Research Imaging Centre (CUBRIC), School of Physics and Astronomy, Cardiff University, Cardiff, United Kingdom\\
$^4$Department of Clinical Sciences, Lund University, Lund, Sweden\\
$^5$Harvard Medical School, Boston, MA, United States\\
$^6$Cardiff University Brain Research Imaging Centre (CUBRIC), School of Psychology, Cardiff University, Cardiff, United Kingdom
}

\date{\today}

\begin{abstract}
Characterizing the properties and orientations of sub-voxel fiber populations, although essential to study white-matter architecture, microstructure and connectivity, remains one of the main challenges faced by the MRI microstructure community. While some progress has been made in overcoming this challenge using models, signal representations and tractography algorithms, these approaches are ultimately limited by their key assumptions or by the lack of specificity of the diffusion signal alone. In order to alleviate these limitations, we combine diffusion-relaxation MR acquisitions incorporating tensor-valued diffusion encoding, Monte-Carlo signal inversions that extract non-parametric intra-voxel distributions of diffusion tensors and relaxation rates, and density-based clustering techniques. This new approach, called "Monte-Carlo density-peak clustering" (MC-DPC), first delineates clusters in the diffusion-orientation subspace of the fiber-like diffusion-relaxation components output by Monte-Carlo signal inversions and then draws from the statistical aspect of these inversion algorithms to compute the median and interquartile range of orientation-resolved means of diffusivities and relaxation rates. Evaluating MC-DPC on tensor-valued diffusion-encoded and $T_2$-weighted correlated datasets \textit{in silico} and \textit{in vivo}, we demonstrate its ability to simultaneously capture sub-voxel fiber orientations and cones of uncertainty, and measure fiber-specific diffusion-relaxation properties that are consistent with the known anatomy and existing literature. Straightforwardly translatable to other diffusion-relaxation correlation experiments probing $T_1$ and $T_2^*$, MC-DPC shows potential in tracking bundle-specific patient-control group differences and longitudinal microstructural changes, enabling new tools for microstructure-informed tractography, and mapping tract-specific myelination states.
\end{abstract}

\maketitle

\footnotetext[1]{\textbf{Abbreviations used:} MRI, magnetic resonance imaging; dMRI, diffusion MRI; HARDI, high angular resolution diffusion imaging; DIAMOND, distribution of anisotropic microstructural environments in diffusion-compartment imaging; ODF, orientation distribution function; DPC, density-peak clustering; MC-DPC, Monte-Carlo density-peak clustering; SH, spherical harmonic; SNR, signal-to-noise ratio.}


\section{Introduction}

Since its debut 15 years ago, the field of connectomics \citep{Hagmann:2005,Sporns:2005} has underlined the importance of white-matter architecture in understanding neurodevelopment, neuroplasticity and neurodegeneration.~\citep{Fornito:2015,Zuo:2017} Enhanced sensitivity to this architecture has been gained with diffusion MRI (dMRI), which captures the translational motion of water molecules diffusing in biological tissue.~\citep{LeBihan:1990, LeBihan:1992, Basser:1994, Mattiello:1994, Mattiello:1997, Jones:2010} Two main research interests have sprung from this non-invasive imaging modality: \textit{microstructure}, which aims at characterizing the content of typical dMRI voxels, whose cubic-millimeter volume encompasses multiple cell types and the extra-cellular space,~\citep{Norris:2001,Sehy:2002,Minati:2007,Mulkern:2009} and \textit{tractography},~\citep{Mori:1999,Basser:2000,Morris:2008,Reisert:2011,Christiaens:2015,Neher:2017,Konopleva:2018,Poulin:2019} which focuses on reconstructing white-matter pathways using the orientation information contained in the diffusion signal. On the one hand, the applications of microstructural imaging have motivated the design of an array of techniques dedicated to mitigating the poor specificity of the diffusion signal. Traditionally, these include various models and signal representations relating the diffusion signal to the voxel content.~\citep{Yablonskiy:2003,Assaf:2004, Assaf_CHARMED:2005,Jbabdi:2012,Zhang_NODDI:2012,Scherrer_DIAMOND:2016,Lampinen_CODIVIDE:2017,Novikov_WMSM:2018} More recently, `tensor-valued' diffusion encoding gradient waveforms \citep{Eriksson:2013,Westin:2014,Eriksson:2015,Westin:2016,Topgaard:2017,Topgaard_dim_rand_walks:2019,Tax_dot:2020} have enhanced the specificity of the data itself by targeting specific features of the intra-voxel diffusion profile. Such measurements use the trace $\mathit{b}$ (size), normalized anisotropy $\mathit{b}_\Delta\in [-0.5,1]$ (shape) and orientation $(\Theta,\Phi)$ of an encoding tensor $\mathbf{b}$.~\citep{Basser:1994,Mattiello:1994,Mattiello:1997} On the other hand, the tractography field has sparked the advent of methods dedicated to resolving sub-voxel white-matter fiber crossings, as fiber crossings occur in 60 to 90\% of voxels in a typical whole-brain imaging experiment.~\citep{Jeurissen:2013} These methods include high angular resolution diffusion imaging (HARDI),~\citep{Tuch:2002} q-ball imaging~\citep{Tuch:2003,Tuch:2004} and spherical deconvolution.~\citep{Tournier:2004, Tournier:2007, Jeurissen:2014} However, mapping the human connectome based on diffusion-orientation information alone remains a challenge to this day,~\citep{Jones_connectivity:2010, Maier-Hein:2017, Schilling:2019, Sotiropoulos_Zalesky:2019} which indicates that additional fiber-specific information is required.

To our knowledge, very few techniques allow for the individual characterization of multiple sub-voxel fiber populations in terms of non-diffusion-orientation quantities, \textit{e.g.} diffusivities and relaxation rates. Notable recent examples include:
\begin{itemize}
\item the composite hindered and restricted model of diffusion (CHARMED),~\citep{Assaf:2004,Assaf_CHARMED:2005} which can capture a hindered extra-axonal compartment described by an effective diffusion tensor, and multiple intra-axonal compartments individually described by a restricted model of diffusion within cylinders.
\item techniques based on orientation distribution functions (ODFs) obtained \textit{via} spherical deconvolution, which provide orientation-resolved measures related to the amplitudes of the ODF lobes.~\citep{Raffelt:2012,dellAcqua:2013}
\item the distribution of anisotropic microstructural environments in diffusion-compartment imaging (DIAMOND) model,~\citep{Scherrer_DIAMOND:2016,Scherrer_aDIAMOND:2017,Reymbaut_arxiv:2020} which separately characterizes sub-voxel crossing anisotropic diffusion compartments using parametric distributions of diffusion tensors.
\item tract-specific $\mathit{T}_1$ mapping,~\citep{deSantis_T1:2016,Andrews_ISMRM:2019} combining inversion recovery and dMRI \citep{deSantis_relaxometry:2016,Hutter:2018,Park_ISMRM:2018} which detects differences in $\mathit{T}_1$ values between white-matter bundles in the healthy brain.
\item the convex optimization modeling for microstructure-informed tractography enhanced by $T_2$ (COMMIT-T2),~\citep{Barakovic_thesis:2019} that estimates a distinct $T_2$ value for each streamline in a tractogram.
\item the joint relaxation-diffusion imaging moments of Ref.~\onlinecite{Ning:2020}, which enable the definition of fast and slow diffusivity/relaxation filters that modulate ODFs according to the diffusivity/relaxation properties of the underlying fiber populations. 
\end{itemize}
However, CHARMED and DIAMOND rely on assumptions regarding the nature of the voxel content and implicitly consider single voxel-averaged longitudinal and transverse relaxation times $T_1$ and $T_2$. Moreover, the works of Refs.~\onlinecite{Raffelt:2012,dellAcqua:2013} rely on assumptions related to tissue-specific response functions that may disagree with the underlying tissue microstructure,~\citep{Parker:2013,Tax:2014} and also implicitly consider single voxel-averaged $T_1$ and $T_2$ values. Finally, while the works of Refs.~\onlinecite{deSantis_T1:2016,Andrews_ISMRM:2019} account for fiber-specific $\mathit{T}_1$ values but involve voxel-content assumptions, the work of Ref.~\onlinecite{Ning:2020} accounts for fiber-specific $\mathit{T}_2$ values but is based on a truncated cumulant expansion that may not properly describe all tissue configurations.~\citep{Reymbaut_accuracy_precision:2020}
In particular, voxel-content assumptions may render a given technique unreliable against partial voluming, more broadly referred to as \textit{tissue heterogeneity}, especially in the presence of pathological tissue. They would also preclude the possibility of identifying differences in $\mathit{T}_1$ and $\mathit{T}_2$ between distinct sub-voxel fiber populations, and characterizing developmental or pathological changes in $\mathit{T}_1$ and $\mathit{T}_2$ within a given sub-voxel fiber population. In particular, tract-specific $\mathit{T}_1$ values enable the evaluation of changes in bundle-specific myelin contents, which are relevant to the study of neurodevelopment, plasticity, aging and neurological disorders.~\citep{van_den_Heuvel:2010,Caeyenberghs:2016,Mancini:2018} Finally, resolving relaxation times along white-matter fibers is of interest in the study of the angular dependence of relaxation times in white matter with respect to the main MRI magnetic field $\mathbf{B}_0$, as observed for $\mathit{T}_1$ \citep{Henkelman:1994,Knight:2018} and $\mathit{T}_2^{(*)}$.~\citep{Henkelman:1994, Bender_Klose:2010, Lee:2011, Rudko:2014, Knight:2015, Gil:2016, McKinnon:2019}


In the context of diffusion-relaxation MRI, a common description of the sub-voxel composition of heterogeneous tissues translates as a distribution $\mathcal{P}(\mathbf{D},\mathit{R}_2,\mathit{R}_1)$ of apparent diffusion tensors~\citep{Jian:2007} $\mathbf{D}$ and apparent relaxation rates $\mathit{R}_2 = 1/\mathit{T}_2$ and $\mathit{R}_1 = 1/\mathit{T}_1$.~\citep{deAlmeidaMartins_Topgaard:2018} The validity of this description is discussed in Appendix~\ref{App_DTD_description}. While the main features of $\mathcal{P}(\mathbf{D},\mathit{R}_2,\mathit{R}_1)$ may be estimated parametrically \textit{via} modeling and/or functional assumptions,~\citep{Alexander:2001,Tuch:2002,Yablonskiy:2003,Kroenke:2004,Jian:2007,Jespersen:2007,Leow:2009,Pasternak:2009,Wang:2011,Fieremans:2011,Zhang_NODDI:2012, Roding:2012, Lasic:2014, Jelescu:2016,Kaden:2016,Westin:2016,Scherrer_DIAMOND:2016,Scherrer_aDIAMOND:2017,Lampinen_CODIVIDE:2017,Reisert:2017,Novikov_on_modeling:2018,Novikov_WMSM:2018,Rensonnet:2018,Reymbaut_arxiv:2020} non-parametric methods evaluating these features have also been developed.~\citep{deAlmeidaMartins_Topgaard:2016,deAlmeidaMartins_Topgaard:2018,Topgaard:2019,deAlmeidaMartins:2020} Such methods, dubbed "Monte-Carlo inversions",~\citep{Prange:2009} estimate $\mathcal{P}(\mathbf{D},\mathit{R}_2,\mathit{R}_1)$ as a discrete set of components $\{( \mathbf{D}_\mathit{n},\mathit{R}_{2,\mathit{n}},\mathit{R}_{1,\mathit{n}}, \mathit{w}_\mathit{n}) \}$, indexed by an integer $\mathit{n}$ and weighted by a discrete probability distribution $\{\mathit{w}_\mathit{n} \}$, without the use of regularization.~\citep{deAlmeidaMartins_Topgaard:2018} Indeed, the estimation of $\mathcal{P}(\mathbf{D},\mathit{R}_2,\mathit{R}_1)$ is instead refined by performing bootstrapping with replacement \citep{de_Kort:2014} on the set of acquired signals, estimating $\{ (\mathbf{D}_\mathit{n},\mathit{R}_{2,\mathit{n}},\mathit{R}_{1,\mathit{n}},\mathit{w}_\mathit{n}) \}$ for each bootstrap solution and computing statistics across all bootstrap solutions.


In this work, we propose a novel approach that estimates the median and interquartile range of orientation-specific metrics across the bootstrap solutions of any non-parametric Monte-Carlo inversion: Monte-Carlo density-peak clustering (MC-DPC). Based on the density-peak clustering (DPC) technique of Ref.~\onlinecite{Rodriguez_Laio:2014}, MC-DPC first delineates clusters in the diffusion-orientation subspace of the all-bootstrap ensemble of components $\{ (\mathbf{D}_\mathit{n},\mathit{R}_{2,\mathit{n}},\mathit{R}_{1,\mathit{n}}, \mathit{w}_\mathit{n} )\}$ with diffusion anisotropy larger than a chosen threshold value. Interpreting the resulting clusters as orientational regions of interest associated with specific sub-voxel fiber populations, it then computes the medians and interquartile ranges of various fiber-specific metrics across bootstrap solutions. As a proof of concept, we evaluate MC-DPC on diffusion- and $\mathit{T}_2$-weighted correlated \textit{in vivo} and \textit{in silico} datasets. We retrieve sub-voxel fiber orientations that are consistent with the known anatomy, alongside orientation-specific means of the isotropic diffusivity $\mathit{D}_\mathrm{iso}=\mathrm{Tr}(\mathbf{D})/3$, squared normalized diffusion anisotropy $\mathit{D}_\Delta^2$ and transverse relaxation rate $\mathit{R}_2$.

\section{Material and methods}
\label{Sec_Methods}

In Section~\ref{Sec_in_vivo_data}, we first describe the methods used to acquire our \textit{in vivo} diffusion-relaxation dataset. We then detail the inner workings of the Monte-Carlo inversion (Sections~\ref{Sec_signal_fitting}, \ref{Sec_statistical_descriptors} and \ref{Sec_ODF_peaks}) and lay down the theory of DPC (Section~\ref{Sec_DPC}) and MC-DPC (Section~\ref{Sec_MC-DPC}). We also explain in Section~\ref{Sec_in_silico_data} how our \textit{in silico} evaluation of MC-DPC was performed. 


\subsection{In vivo human brain data}
\label{Sec_in_vivo_data}

\begin{figure*}[ht!]
\begin{center}
\includegraphics[width=35pc]{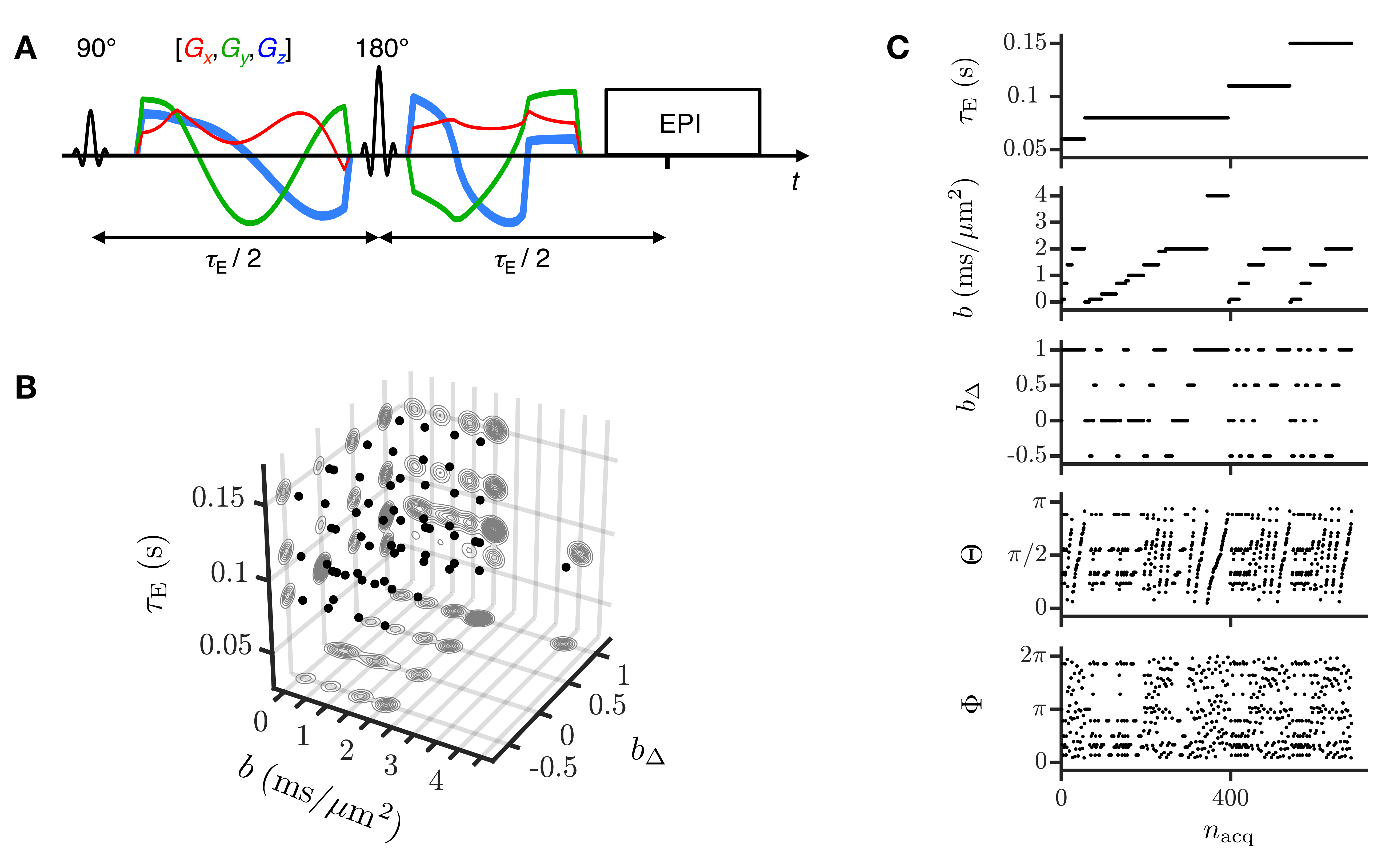}
\caption{Visualization of our acquisition scheme, identical to the one used in Ref.~\onlinecite{deAlmeidaMartins_ISMRM:2019}. (A) Spin-echo sequence with EPI readout customized for free-waveform encoding and variable echo times $\tau_\mathrm{E}$. (B) 5D grid-like acquisition scheme where black points indicate the acquisition points in the 3D subspace of echo time $\tau_\mathrm{E}$, b-tensor size $b$ and b-tensor shape $b_\Delta \in [-0.5, 1]$. The number of b-tensor orientations $(\Theta,\Phi)$ for each point is illustrated by the projected contours. (C) Acquisition parameters as a function of sorted acquisition point index $n_\mathrm{acq}$.}
\label{Figure_acq}
\end{center}
\end{figure*}

The study was approved by the Cardiff University School of Psychology Ethics Committee and written informed consent was obtained from the participant. This healthy volunteer was scanned on a 3T Siemens MAGNETOM Prisma equipped with a 32-channel receiver head-coil, using a prototype spin-echo sequence with EPI readout, TR=4 s, FOV=234x234x60 mm$^3$, voxel-size=3x3x3 mm$^3$, partial-Fourier=6/8 and iPAT=2 (GRAPPA),
customized for tensor-valued diffusion encoding \citep{Lasic:2014,Szczepankiewicz_DIVIDE:2019} and variable echo time $\tau_\mathrm{E}$ (Figure~\ref{Figure_acq}.A). Tensor-valued diffusion encoding was performed with numerically optimized \citep{Sjolund:2015} and Maxwell-compensated \citep{Szczepankiewicz_Maxwell:2019} waveforms. We also attempted to match their respective frequency contents.~\citep{Lundell:2019} Constraining the diffusion tensors in our distributions to be axisymmetric,~\citep{Eriksson:2015} the dimensions of our 45-minute 686-point acquisition scheme (Figures~\ref{Figure_acq}.B-\ref{Figure_acq}.C), identical to the one used in Ref.~\onlinecite{deAlmeidaMartins_ISMRM:2019}, match those of $\mathcal{P}(\mathbf{D},\mathit{R}_2)$. Images were motion- and eddy-topup corrected using the procedures described in Refs.~\onlinecite{Andersson_topup:2003,Nilsson:2015}. The signal-to-noise ratio (SNR) of this dataset was estimated to be 90 across voxels of the corona radiata by computing the mean-to-standard-deviation ratio of the spherically encoded diffusion signal at $b=0.3$ ms/{\textmu}m$^2$ and $\tau_\mathrm{E}=80$ ms (see Supplemental Material of Ref.~\onlinecite{Szczepankiewicz_DIVIDE:2019}).

\subsection{Monte-Carlo signal inversion}
\label{Sec_MC_inversion}

\subsubsection{Signal fitting and bootstrapping}
\label{Sec_signal_fitting}

Laplace inversion of the diffusion-relaxation signal typically relies on regularization in order to guide the search for a suitable solution to the inversion problem.~\citep{Provencher:1982,Kroeker:1986,Whittall:1989,Mitchell:2012} However, the memory requirements of these algorithms make them impractical for high dimensions. Besides, the choice of regularization term inherently influences the characteristics of the solution, potentially causing a mismatch between these characteristics and those of the underlying microstructure. To circumvent these problems, we instead used the Monte-Carlo inversion algorithm described in Ref.~\onlinecite{deAlmeidaMartins_Topgaard:2018} to analyze the diffusion-$R_2$ dataset described in Section~\ref{Sec_in_vivo_data}. This algorithm explores the $(\mathbf{D},R_2)$ solution space using a stochastic iterative approach that does not rely on regularization nor on assumptions regarding the nature of the voxel content.

Considering axisymmetric diffusion tensors, parametrized by the axial diffusivity $\mathit{D}_\parallel = \mathit{D}_\mathrm{iso} (1+2D_\Delta)$, radial diffusivity $\mathit{D}_\perp=\mathit{D}_\mathrm{iso} (1-D_\Delta)$ and orientation $(\theta, \phi)$, with the isotropic diffusivity $\mathit{D}_\mathrm{iso}$ and normalized anisotropy $\mathit{D}_\Delta \in [-0.5, 1]$,~\citep{Haeberlen:1976,Eriksson:2015} the Monte-Carlo inversion technique of Ref.~\onlinecite{deAlmeidaMartins_Topgaard:2018} retrieves non-parametric 5D intra-voxel distributions $\mathcal{P}(\mathbf{D},\mathit{R}_2) \equiv \mathcal{P}(\mathit{D}_\parallel,\mathit{D}_\perp,\theta,\phi, \mathit{R}_2)$ from diffusion-$R_2$ datasets. Given that the $\mathit{T}_2$-weighting of the dataset detailed in Section~\ref{Sec_in_vivo_data} is provided by a variable echo time $\tau_\mathrm{E}$, the inversion algorithm inverts the signal equation
\begin{widetext}
\begin{equation}
\frac{\mathcal{S}(\mathbf{b},\tau_\mathrm{E})}{\mathcal{S}_0} = \int_0^{+\infty}\int_{\mathrm{Sym}^+(3)} \mathcal{P}(\mathbf{D},\mathit{R}_2)\,\exp(-\mathbf{b}:\mathbf{D})\,\exp(-\tau_\mathrm{E}\mathit{R}_2)\, \mathrm{d}\mathbf{D}\,\mathrm{d}\mathit{R}_2\,,
\end{equation}
\end{widetext}
where $\mathbf{b}$ is the diffusion-encoding tensor from tensor-valued diffusion encoding,~\citep{Eriksson:2013,Westin:2014,Eriksson:2015,Westin:2016,Topgaard:2017,Topgaard_dim_rand_walks:2019} $\mathcal{S}_0 = \mathcal{S}(\mathbf{b}=\mathbf{0},\mathrm{\tau}_\mathrm{E}\to 0)$, $\mathrm{Sym}^{+}(3)$ denotes the space of symmetric positive-semidefinite 3$\times$3 tensors, and ":" is the Frobenius inner product. For axisymmetric b-tensors, this generalized scalar product writes \citep{Eriksson:2015}
\begin{equation}
\mathbf{b}:\mathbf{D} = bD_\mathrm{iso}[1 + 2b_\Delta D_\Delta P_2(\cos\beta)] \, ,
\end{equation}
where $\mathit{P}_2(\mathit{x}) = (3\mathit{x}^2-1)/2$ is the second Legendre polynomial and $\cos\beta = \cos\Theta \cos\theta + \sin\Theta\sin\theta\cos(\Phi - \phi)$ is the cosine of the shortest angle $\beta$ between the main axis $(\Theta,\Phi)$ of $\mathbf{b}$ and the main axis $(\theta,\phi)$ of $\mathbf{D}$. 
For computational purposes, this equation is discretized for each acquired signal $\mathcal{S}_m$ as a finite sum of $\mathit{N}$ components $(\mathbf{D}_\mathit{n}, \mathit{R}_{2,\mathit{n}})\equiv (\mathit{D}_{\parallel,\mathit{n}}, \mathit{D}_{\perp,\mathit{n}} , \theta_\mathit{n}, \phi_\mathit{n}, \mathit{R}_{2,\mathit{n}})$:
\begin{equation}
\mathcal{S}_\mathit{m} = \sum_{\mathit{n}=1}^\mathit{N} \mathit{w}_\mathit{n}\,\exp(-\mathbf{b}_\mathit{m}:\mathbf{D}_\mathit{n})\,\exp(-\tau_{\mathrm{E},\mathit{m}}\mathit{R}_{2,\mathit{n}})\, ,
\label{Eq_signal_discretized}
\end{equation}
where $\mathit{w}_\mathit{n}$ is the weight of a given component, normalized so that $\sum_{\mathit{n}=1}^\mathit{N} \mathit{w}_\mathit{n} = \mathcal{S}_0$. A short-hand notation of Equation~\ref{Eq_signal_discretized} reads
\begin{equation}
\mathbf{S} = \mathbf{K}\mathbf{w}\, ,
\end{equation}
where $\mathbf{S}$ is the column vector containing the acquired signals $\mathcal{S}_m$, $\mathbf{K}$ is the inversion kernel matrix containing the signal decays and $\mathbf{w}$ is the column vector containing the weights $\mathit{w}_n$ of the components $(\mathbf{D}_n, \mathit{R}_{2,n})$.
The Monte-Carlo inversion algorithm randomly samples such components and estimates the associated vector $\mathbf{w}$ quantifying their propensity to fit the acquired signals \textit{via} non-negative least-squares fitting:~\citep{Lawson_book:1974,Whittall:1989,English:1991,Venkataramanan:2002,Mitchell:2012}
\begin{equation}
\mathbf{w} = \underset{\mathbf{w}^\prime\geq 0}{\mathrm{argmin}}\left[\Vert\mathbf{S}-\mathbf{K}\mathbf{w}^\prime\Vert_2^2\right] ,
\end{equation}
where $\Vert\cdot\Vert_2$ denotes the L$_2$ norm. This process is repeated iteratively following a quasi-genetic filtering detailed in Refs.~\onlinecite{deAlmeidaMartins_Topgaard:2016, deAlmeidaMartins_Topgaard:2018, Topgaard:2019,deAlmeidaMartins:2020}.
%
Embracing the inherent ill-conditioning of Laplace inversion problems, we performed bootstrapping with replacement \citep{de_Kort:2014} on the data and estimated for each voxel an ensemble of $\mathit{N}_\mathrm{b}=96$ plausible sets of components, also called "bootstrap solutions", each denoted by $\{(\mathit{D}_{\parallel,\mathit{n}}, \mathit{D}_{\perp,\mathit{n}} , \theta_\mathit{n}, \phi_\mathit{n}, \mathit{R}_{2,\mathit{n}}, \mathit{w}_\mathit{n})\}_{1\leq \mathit{n}\leq \mathit{N}=20}$. We then computed statistical descriptors of $\mathcal{P}(\mathbf{D},\mathit{R}_2)$ for each bootstrap solution and calculated the median of each statistical descriptor across bootstrap solutions (see Section~\ref{Sec_statistical_descriptors}).

\subsubsection{Statistical descriptors}
\label{Sec_statistical_descriptors}

The final solution of the Monte-Carlo inversion algorithm, $\mathcal{P}(\mathbf{D},\mathit{R}_2)$, can be understood as the median of the solutions obtained for each bootstrap solution, $\mathcal{P}_{\mathit{n}_\mathrm{b}}(\mathbf{D},\mathit{R}_2)$ with $1\leq \mathit{n}_\mathrm{b} \leq \mathit{N}_\mathrm{b}$. To quantify the main features of this final solution, one computes for instance the medians over bootstrap solutions of the per-bootstrap mean diffusivity, mean squared normalized anisotropy, and mean transverse relaxation rate:
\begin{equation}
\underset{n_\mathrm{b}}{\mathrm{Med}}\,(\mathrm{E}[\chi]_{n_\mathrm{b}})\, ,
\end{equation}
with $\chi=\mathit{D}_\mathrm{iso}, D_\Delta^2, R_2$, respectively. Here, the median operator "$\mathrm{Med}$" acts across bootstrap solutions and $\mathrm{E}[\,\cdot\,]_{\mathit{n}_\mathrm{b}}$ denotes the per-bootstrap average over the solution space retrieved within the bootstrap solution $\mathit{n}_\mathrm{b}$. This idea of using means as distribution descriptors and quartile-based statistical objects such as medians/interquartile ranges as averages/dispersions across bootstrap solutions is consistent with previous works.~\citep{Reymbaut_accuracy_precision:2020,deAlmeidaMartins:2020}

\subsubsection{Binning, orientation distribution functions and peaks}
\label{Sec_ODF_peaks}

By design, the Monte-Carlo inversion algorithm progressively builds up the sought-for intra-voxel distribution $\mathcal{P}(\mathbf{D},\mathit{R}_2)$ as a discrete non-parametric weighted sum of components. This implies that tissue-specific statistical descriptors can be extracted by subdividing the 5D configuration space of $\mathcal{P}(\mathbf{D},\mathit{R}_2)$ into multiple bins.~\citep{Topgaard:2019, deAlmeidaMartins:2020} Such bins include the "thin", "thick" and "big" bins introduced in Ref.~\onlinecite{deAlmeidaMartins:2020}, aiming to isolate the signal contributions from white matter, grey matter and cerebrospinal fluid, respectively. In particular, the "thin" bin isolates components of high normalized diffusion anisotropy $\mathit{D}_\Delta \geq 0.5$. Orientation distribution functions (ODFs) can be defined from these thin-bin components using the procedure detailed in Refs.~\onlinecite{deAlmeidaMartins_ISMRM:2019,deAlmeidaMartins_thesis:2020}. For each bootstrap solution $\mathit{n}_\mathrm{b}$ (with $1\leq\mathit{n}_\mathrm{b}\leq \mathit{N}_\mathrm{b}$), we considered the voxel-wise discrete ensemble of components belonging to the thin bin, 
\begin{equation}
\mathcal{E}^\mathrm{thin}_{n_\mathrm{b}} = \{(\mathit{D}_{\parallel,\mathit{i}}, \mathit{D}_{\perp,\mathit{i}} , \theta_\mathit{i}, \phi_\mathit{i}, \mathit{R}_{2,\mathit{i}}, \mathit{w}_\mathit{i})\}_{\mathit{n}_\mathrm{b},\;\mathit{i}\in \{\text{thin bin}\}}\, ,
\label{Eq_per_bootstrap_solutions}
\end{equation}
and computed an ODF $\mathit{P}_{\mathit{n}_\mathrm{b}}(\theta_\text{mesh},\phi_\text{mesh})$ on a spherical mesh $\{(\theta_\text{mesh},\phi_\text{mesh})\}$ by convolving the discrete set of components of Equation~\ref{Eq_per_bootstrap_solutions} with a Watson distribution kernel \citep{Watson:1965,Mardia_Jupp:2009} as
\begin{equation}
\mathit{P}_{\mathit{n}_\mathrm{b}}(\theta_\text{mesh},\phi_\text{mesh}) = \sum_{\mathit{i}\in \mathcal{E}^\mathrm{thin}_{n_\mathrm{b}}} \mathit{w}_\mathit{i}\,\exp (\kappa\, [\mathbf{u}_\mathit{i}\cdot\bm{\mu}(\theta_\text{mesh},\phi_\text{mesh})]^2)\, ,
\label{Eq_per_bootstrap_ODF}
\end{equation}
where $\mathbf{u}_\mathit{i} \equiv (\theta_i,\phi_i)$ is the unit vector giving the orientation of component $\mathit{i}$, $\bm{\mu}(\theta_\text{mesh},\phi_\text{mesh})\equiv (\theta_\text{mesh},\phi_\text{mesh})$ is the unit vector associated with a spherical-mesh point, "$\cdot$" denotes the scalar product, and $\kappa$ is the concentration parameter that regulates the amount of orientation dispersion around $\mathbf{u}_\mathit{i}$ in the Watson kernel. A final voxel-wise ODF $\mathit{P}(\theta_\text{mesh},\phi_\text{mesh})$ was calculated as the median of the per-bootstrap ODFs:
\begin{equation}
\mathit{P}(\theta_\text{mesh},\phi_\text{mesh}) = \underset{n_\mathrm{b}}{\mathrm{Med}}\left( \mathit{P}_{\mathit{n}_\mathrm{b}}(\theta_\text{mesh},\phi_\text{mesh}) \right) \, .
\end{equation}
The convolution in Equation~\ref{Eq_per_bootstrap_ODF} can be understood as giving an effective `orientational uncertainty' to each component $\mathbf{u}_\mathit{i}$ \textit{via} $\kappa$, finally estimating this collection of $\mathbf{u}_\mathit{i}$-centered Watson distributions at the spherical-mesh points. In other words, the purpose of the Watson kernel is to smoothly map the discrete set of components of Equation~\ref{Eq_per_bootstrap_solutions} onto the nearest nodes on the spherical mesh $\{(\theta_\text{mesh},\phi_\text{mesh})\}$. In this work, we considered a $1000$-point uniform spherical mesh and set the concentration parameter to $\kappa=14.9$, following the rationale detailed in Appendix~\ref{App_kappa_ODF}.

The set of diffusion-relaxation metrics $\{(\mathit{D}_{\parallel,\mathit{i}}, \mathit{D}_{\perp,\mathit{i}}, \mathit{R}_{2,\mathit{i}}\}_{\mathit{n}_\mathrm{b},\;\mathit{i}\in \{\text{thin bin}\}}$ can also be mapped onto this spherical mesh by computing the median across bootstrap solutions of the per-bootstrap average $\mathrm{E}[\chi]_{\mathit{n}_\mathrm{b}}$ projected onto our estimated per-bootstrap ODFs in $(\theta_\text{mesh},\phi_\text{mesh})$, \textit{i.e.} $\hat{\mathrm{E}}[\chi]_{\mathit{n}_\mathrm{b}}(\theta_\text{mesh},\phi_\text{mesh})$:~\citep{deAlmeidaMartins_ISMRM:2019,deAlmeidaMartins_thesis:2020}
\begin{widetext}
\begin{equation}
\underset{n_\mathrm{b}}{\mathrm{Med}}\left(\hat{\mathrm{E}}[\chi]_{\mathit{n}_\mathrm{b}}(\theta_\text{mesh},\phi_\text{mesh})\right) = \underset{n_\mathrm{b}}{\mathrm{Med}}\left(\frac{1}{\mathit{P}_{\mathit{n}_\mathrm{b}}(\theta_\text{mesh},\phi_\text{mesh})} \sum_{\mathit{i}\in \mathcal{E}^\mathrm{thin}_{n_\mathrm{b}}} \mathit{w}_\mathit{i}\,\chi_\mathit{i}\,\exp (\kappa\, [\mathbf{u}_\mathit{i}\cdot\bm{\mu}(\theta_\text{mesh},\phi_\text{mesh})]^2) \right) \equiv \hat{\mathrm{E}}[\chi] \, ,
\label{Eq_ODF_metrics}
\end{equation}
\end{widetext}
where $\chi \equiv \mathit{D}_\mathrm{iso}, \mathit{D}_\Delta^2, \mathit{R}_2, \mathit{T}_2 = 1/\mathit{R}_2$. The short-hand notation "$\hat{\mathrm{E}}[\chi]$" is now retained in Equation~\ref{Eq_ODF_metrics} for simplicity. Note that the ODF metrics associated with $T_2$ and $R_2$ in Equation~\ref{Eq_ODF_metrics} are computed separately, as both quantities are commonly found in the MRI literature. Moreover, the average of an inverse does not equal the inverse of an average.

\begin{figure*}[ht!]
\begin{center}
\includegraphics[width=35pc]{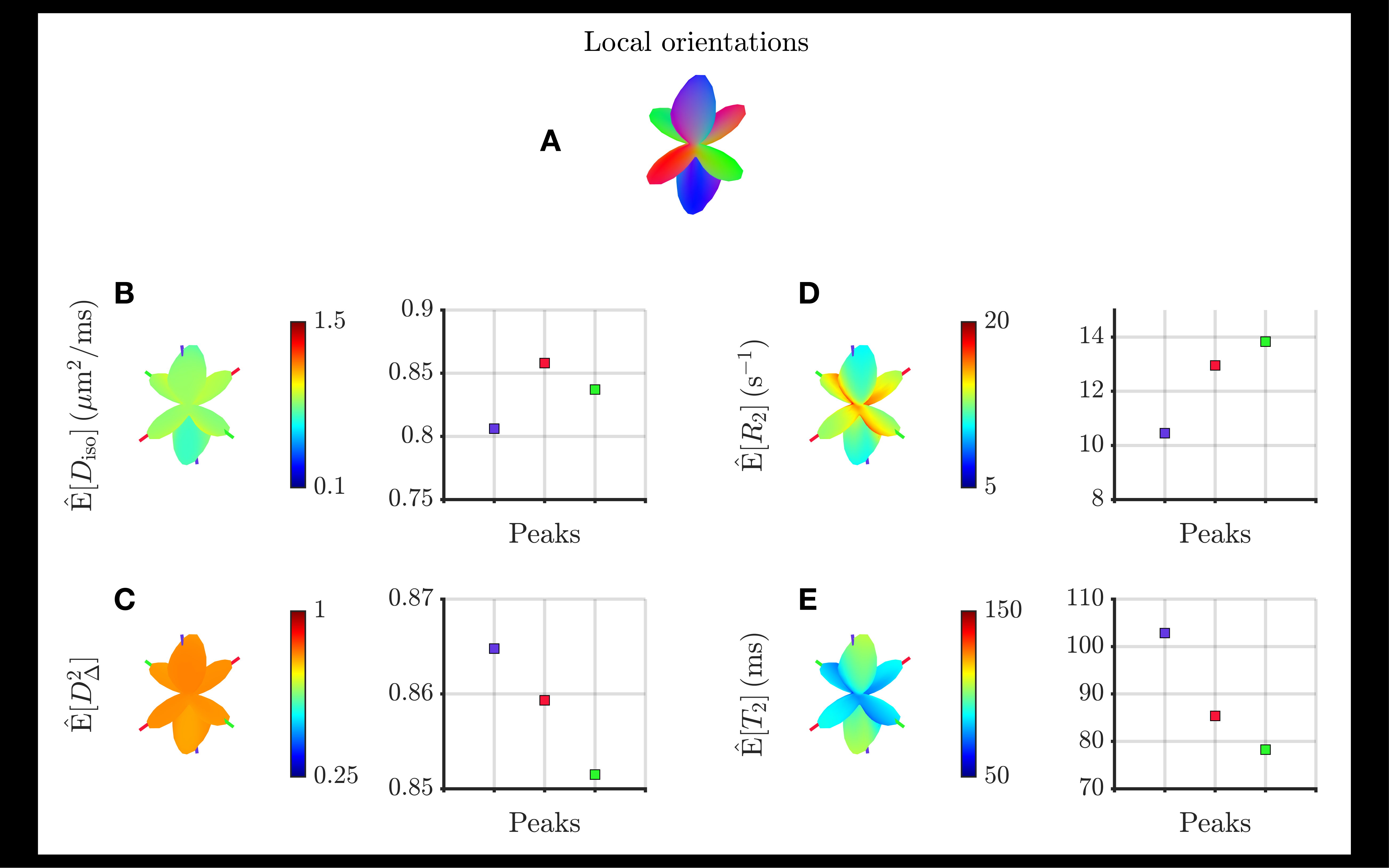}
\caption{Typical ODF and associated peaks and peak metrics in the \textit{in vivo} three-way crossing detailed in Figure~\ref{Figure_CR}. The color mapped onto the ODF either codes for local orientation according to $\mathrm{[red, green, blue]}\equiv [\vert x \vert, \vert y \vert, \vert z \vert]/\mathrm{max}([\vert x \vert, \vert y \vert, \vert z \vert])$ (A) or for the ODF metrics Equation~\ref{Eq_ODF_metrics} $\hat{\mathrm{E}}[D_\mathrm{iso}]$ (B), $\hat{\mathrm{E}}[D_\Delta^2]$ (C), $\hat{\mathrm{E}}[R_2]$ (D) and $\hat{\mathrm{E}}[T_2]$ (E), with corresponding color bars. Grid plots indicate the value of the ODF metrics at the ODF peaks represented by orientation-colored sticks through the ODFs (B,C,D,E). The colors of the ODF-peak metric points match those of their respective ODF peaks.}
\label{Figure_ODF_pedagogic}
\end{center}
\end{figure*}

ODFs are practical to define "peaks", \textit{i.e.} local maxima of $\mathit{P}(\theta_\text{mesh},\phi_\text{mesh})$ that tease apart main diffusion orientations in the intra-voxel diffusion profile. Such local maxima were found on the parts of the discrete ODF such that $\mathit{P}(\theta_\text{mesh},\phi_\text{mesh})/\mathrm{max}(\mathit{P})\geq 0.1$, keeping up to a maximum of four peaks per voxel. One can then assign metrics to these peaks using Equation~\ref{Eq_ODF_metrics} to estimate $\hat{\mathrm{E}}[\chi]$ for each peak orientation. An \textit{in vivo} example of an ODF with its associated peaks and peak metrics is provided in Figure~\ref{Figure_ODF_pedagogic}. 
Whereas the ODFs derived from the output of the Monte-Carlo signal inversion offer a convenient representation of orientation-specific features,~\citep{deAlmeidaMartins_thesis:2020} they do not allow for a proper quantification of the median value and precision of these features across bootstrap solutions. 

\subsection{Density-peak clustering (DPC)}
\label{Sec_DPC}

Clustering consists of grouping a set of data objects into multiple subsets called \textit{clusters}, so that objects belonging to the same cluster share more \textit{similarity}, as defined by clustering criteria, than objects from different clusters. Appendix~\ref{App_clustering_techniques} provides a brief review of various clustering strategies that have been investigated \citep{Xu_Wunsch:2005} and explains why the "density-peak clustering" (DPC) technique \citep{Rodriguez_Laio:2014} was chosen as a starting point in this work.

In any given dataset for which a distance $\mathit{d_{ij}}$ between data points $\mathit{i}$ and $\mathit{j}$ is defined, DPC identifies data clusters using two metrics: the density $\rho_\mathit{i}$ at data point $\mathit{i}$ \citep{Cheng:1995}
\begin{equation}
\rho_i = \sum_j \exp(-d_{ij}^2/(2d_\mathrm{cutoff}^2))\, ,
\label{Eq_Gaussian_kernel}
\end{equation}
where $\mathit{d}_\mathrm{cutoff}$ is a cutoff distance, and the delta-distance
\begin{equation}
\delta_i = \underset{j\neq i\, ,\, \rho_j>\rho_i}{\mathrm{min}}\, d_{ij}
\label{Eq_delta_distance}
\end{equation}
separating data point $\mathit{i}$ and the closest point of higher density. While the original work of Ref.~\onlinecite{Rodriguez_Laio:2014} measures the density $\rho_\mathit{i}$ as the number of points distant from point $\mathit{i}$ by less than $\mathit{d}_\mathrm{cutoff}$, it also refers to the continuous Gaussian kernel Equation~\ref{Eq_Gaussian_kernel}, often preferred to a truncated measure of density. The value of $\mathit{d}_\mathrm{cutoff}$ is set arbitrarily and cluster centroids are manually identified in a decision graph on the intuitive basis that cluster centroids are characterized by large values of $\rho$ and $\delta$. Consequently, the number of centroids $\mathit{N}_\mathrm{c}$ is also set manually. After selection of the cluster centroids, DPC iteratively assigns data points to the cluster which contains their nearest neighbors with higher density values. 
If $\mathit{N}_\mathrm{c}>1$, outliers to data clusters can be detected. For each cluster, a border region is identified as the set of data points assigned to this cluster while being within a distance $\mathit{d}_\mathrm{cutoff}$ from data points belonging to other clusters. Denoting by $\rho_\mathrm{b}$ the highest point density computed \textit{via} Equation~\ref{Eq_Gaussian_kernel} within this border region, the points of the cluster whose density is lower than $\rho_\mathrm{b}$ are considered as outliers. 

Multiple improvements have been brought to DPC since the original work of Ref.~\onlinecite{Rodriguez_Laio:2014}. The cutoff distance $\mathit{d}_\mathrm{cutoff}$ can be set automatically using the data-field method described in Ref.~\onlinecite{Wang:2015}. This work draws parallels with statistical physics, using the Gaussian kernel of Equation~\ref{Eq_Gaussian_kernel},
\begin{equation}
\varphi_\sigma(r) = \exp(-r^2/(2\sigma^2))
\end{equation}
with unknown standard deviation $\sigma$, to define the Gaussian data-field potential at data point $i$ \citep{Wang:2011_datafield}
\begin{equation}
\varphi_{\sigma,i} = \sum_j \varphi_\sigma(d_{ij}) = \sum_j \exp(-d_{ij}^2/(2\sigma^2))
\end{equation}
and the data-field entropy
\begin{equation}
H(\sigma) = - \sum_i \frac{\varphi_{\sigma,i}}{Z}\, \ln\left( \frac{\varphi_{\sigma,i}}{Z} \right)
\label{Eq_entropy_DPC}
\end{equation}
with the partition function $Z = \sum_i \varphi_{\sigma,i}$. Denoting by $\sigma_0$ the value that minimizes $H(\sigma)$, the cutoff distance is then set to $d_\mathrm{cutoff} = 3\sigma_0$, so that the Gaussian kernel vanishes significantly over that distance according to the $3\sigma$ rule of Gaussian functions. 
A more subtle outlier detection has also been given in Ref.~\onlinecite{Tao:2017}. Denoting two distinct data clusters by $\mathit{C}_1$ and $\mathit{C}_2$, outliers to these clusters are the data points $\mathit{k}$ satisfying 
\begin{equation}
\rho_k < \underset{\substack{\mathit{i}\in \mathit{C}_1\, ,\, \mathit{j}\in \mathit{C}_2\\\mathit{d_{ij}< d_\mathrm{cutoff}}}}{\mathrm{max}}\! \left( \frac{\rho_i + \rho_j}{2} \right)\, .
\label{Eq_outlier_detection_DPC}
\end{equation}
Finally, assuming that the number of clusters $N_\mathrm{c}$ has been provided by the user, strategies have been designed to automatically select the cluster centroids, either by choosing the $N_\mathrm{c}$ data points with largest delta-distances $\delta$ (see Equation~\ref{Eq_delta_distance}),~\citep{Hou_Pelillo:2016,Yan:2019} or by choosing the $N_\mathrm{c}$ data points with largest values of $\gamma = \rho\delta$ (see Equations~\ref{Eq_Gaussian_kernel} and \ref{Eq_delta_distance}).~\citep{Hou_Pelillo:2016, Wang:2016,Sieranoja_Franti:2019}


\newpage

\begin{widetext}

\subsection{Monte-Carlo density-peak clustering (MC-DPC)}
\label{Sec_MC-DPC}

We designed the Monte-Carlo density-peak clustering (MC-DPC) procedure in two main steps. While the first step consists of using an adapted version of DPC to identify orientational regions of interest from the output of the Monte-Carlo signal inversion, the second step lies in computing statistics of diffusion-relaxation metrics within these regions of interest. The MC-DPC procedure is illustrated in Figure~\ref{Figure_MC-DPC}.

\begin{figure*}[ht!]
\begin{center}
\includegraphics[width=35pc]{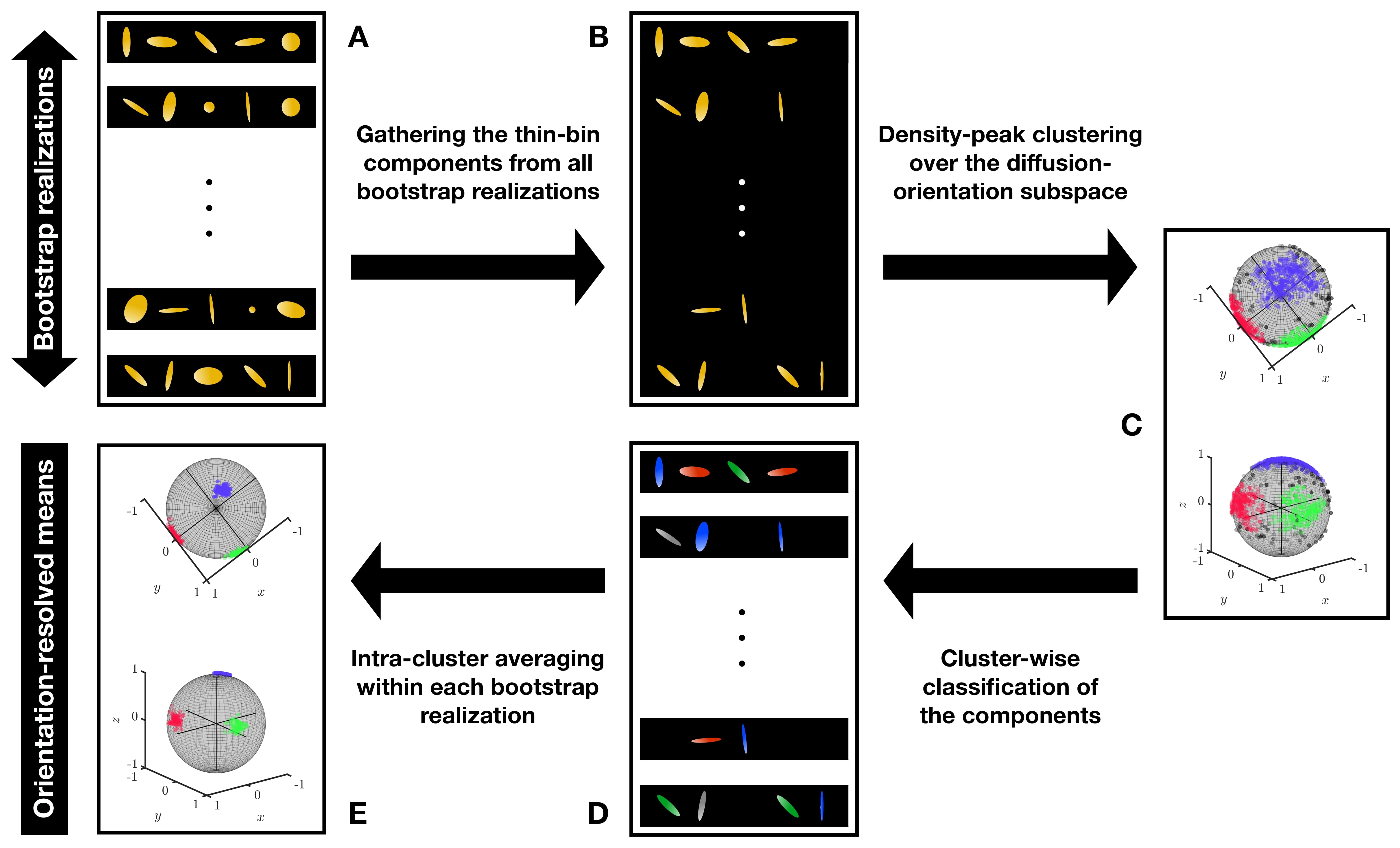}
\caption{Illustration of the MC-DPC procedure in the \textit{in vivo} three-way crossing detailed in Figure~\ref{Figure_CR}. Considering the set of solutions generated by our $N_\mathrm{b}=96$ bootstrap solutions as inputs (A), the thin-bin components of these solutions (B) are clustered in their orientation subspace using an adapted version of DPC (C) so that to classify them in cluster-specific categories (D). Averaging these components independently for each bootstrap solution and cluster-specific category enables to define orientation-resolved means of diffusion-relaxation metrics (E). Panel C presents the orientations of the all-bootstrap thin-bin solutions on the unit sphere. Opacity codes for the weight of each component. Black points are outliers detected by MC-DPC according to Equation~\ref{Eq_outlier_detection_MC_DPC}. Colored points belong to three clusters, with color coding for the averaged orientation of each cluster. Panel E shows the collection of intra-bootstrap intra-cluster mean orientations $(\mathring{\mathrm{E}}[x],\mathring{\mathrm{E}}[y],\mathring{\mathrm{E}}[z])$ Equation~\ref{Eq_orientation_resolved_mean} on the unit sphere. While opacity codes for the weight of the intra-cluster averaged components (see Equation~\ref{Eq_cluster_weight}), color codes for the geometric median orientation of each cluster (see Equation~\ref{Eq_geometric_median}).}
\label{Figure_MC-DPC}
\end{center}
\end{figure*}

First, MC-DPC gathers the ensemble of all per-bootstrap thin-bin solution sets $\mathcal{E}^\mathrm{thin}_{n_\mathrm{b}}$ (see Equation~\ref{Eq_per_bootstrap_solutions}, Figures~\ref{Figure_MC-DPC}.A and \ref{Figure_MC-DPC}.B),
\begin{equation}
\mathcal{E}^\mathrm{thin}_\mathrm{all-bootstrap} = \left\{\mathcal{E}^\mathrm{thin}_{n_\mathrm{b}}\right\}_{1\leq n_\mathrm{b}\leq N_\mathrm{b}} = \left\{\{(\mathit{D}_{\parallel,\mathit{i}}, \mathit{D}_{\perp,\mathit{i}} , \theta_\mathit{i}, \phi_\mathit{i}, \mathit{R}_{2,\mathit{i}}, \mathit{w}_\mathit{i})\}_{n_\mathrm{b},\;\mathit{i}\in \{\text{thin bin}\}}\right\}_{1\leq n_\mathrm{b}\leq N_\mathrm{b}}\, ,
\label{Eq_ensemble_thin_all_bootstrap}
\end{equation}
and delineates $N_\mathrm{c}$ clusters in its orientation subspace $\left\{\{(\theta_\mathit{i}, \phi_\mathit{i})\}_{n_\mathrm{b},\;\mathit{i}\in \{\text{thin bin}\}}\right\}_{1\leq n_\mathrm{b}\leq N_\mathrm{b}}$ using DPC (Figure~\ref{Figure_MC-DPC}.C) with data-point density
\begin{equation}
\rho_i = \sum_{j\in\mathcal{E}^\mathrm{thin}_\mathrm{all-bootstrap}} w_j\,\exp(-d_{ij}^2/(2d_\mathrm{cutoff}^2))
\end{equation}
and distance $\mathit{d_{ij}}$ given by the angular distance
\begin{equation}
\mathit{d_{ij}} = \arccos(\vert \cos\beta_{ij} \vert) = \arccos(\vert \cos\theta_i\cos\theta_j + \sin\theta_i\sin\theta_j\cos(\phi_i - \phi_j) \vert) \equiv d(\mathbf{u}_i, \mathbf{u}_j) \, ,
\label{Eq_distance_MC_DPC}
\end{equation}
where $\beta_{ij}$ is the shortest angle between points $\mathit{i}$ and $\mathit{j}$ on the unit sphere (equivalently, the shortest angle between unit orientations $\mathbf{u}_i\equiv(\theta_i,\phi_i)$ and $\mathbf{u}_j\equiv(\theta_j,\phi_j)$), as given by the spherical law of cosines. The presence of an absolute value $\vert \,\cdot\, \vert$ in this definition of $\mathit{d_{ij}}$ captures the antipodal spatial symmetry of the solution space. The expression of MC-DPC's delta-distance is identical to Equation~\ref{Eq_delta_distance}. The cutoff distance $d_\mathrm{cutoff}$ was set automatically using the aforementioned method of Ref.~\onlinecite{Wang:2015} altered to account for the solution weights so that the Gaussian data-field potential leading to the minimization of the data-field entropy Equation~\ref{Eq_entropy_DPC} writes $\varphi_{\sigma,i} = \sum_j w_j\,\exp(-d_{ij}^2/(2\sigma^2))$. The aforementioned outlier detection of Ref.~\onlinecite{Tao:2017} between two distinct data clusters by $\mathit{C}_1$ and $\mathit{C}_2$ was also altered to account for the solution weights so that the criterion Equation~\ref{Eq_outlier_detection_DPC} now writes
\begin{equation}
\rho_k < \underset{\substack{\mathit{i}\in \mathit{C}_1\, ,\, \mathit{j}\in \mathit{C}_2\\\mathit{d_{ij}< d_\mathrm{cutoff}}}}{\mathrm{max}}\! \left( \frac{\mathit{w}_\mathit{i}\rho_i + \mathit{w}_\mathit{j}\rho_j}{\mathit{w}_\mathit{i} + \mathit{w}_\mathit{j}} \right)\, .
\label{Eq_outlier_detection_MC_DPC}
\end{equation}

Importantly, the optimal number of clusters $N_\mathrm{c}$ was automatically set by a conservative estimate of the number of ODF peaks in a given voxel. This estimate was obtained voxel-wise as the maximum between the number of peaks retrieved from our ODFs according to Section~\ref{Sec_ODF_peaks} and the number of peaks extracted after fitting our ODFs with spherical harmonics (SHs) of maximal SH order of eight using MRtrix.~\citep{Tournier_MRtrix:2019} SH-peak extraction was performed with the same weight threshold as that used in Section~\ref{Sec_ODF_peaks}. The cluster centroids were then chosen as the $N_\mathrm{c}$ data points with largest values of 
\begin{equation}
\gamma = \frac{\rho\delta}{\mathrm{max}(\rho)\,\mathrm{max}(\delta)}\, ,
\end{equation}
following the criterion chosen in previous works \citep{Hou_Pelillo:2016, Wang:2016,Sieranoja_Franti:2019} to identify data points with large values of both density and delta-distance. Note that our conservative initial estimate for $N_\mathrm{c}$ does not necessarily set the final value of $N_\mathrm{c}$, as MC-DPC may filter out clusters whose total weights are below a certain threshold (see Equation~\ref{Eq_MC_DPC_filtering} below).

Assuming that the estimated clusters, resulting from orientational aggregates of the all-bootstrap thin-bin solutions, can be interpreted as orientational regions of interest associated with sub-voxel fiber orientations, we carried on further with MC-DPC and computed orientation-resolved statistics across bootstrap solutions. To do so, we separately classified each per-bootstrap ensemble of thin-bin solutions $\mathcal{E}^\mathrm{thin}_{n_\mathrm{b}}$ Equation~\ref{Eq_per_bootstrap_solutions} into $N_\mathrm{c}$ ensembles $\mathcal{E}^\mathrm{thin}_{n_\mathrm{b},n_\mathrm{c}}$ (with $1\leq n_\mathrm{c}\leq N_\mathrm{c}$), each corresponding to an estimated cluster (Figure~\ref{Figure_MC-DPC}.D). We then averaged the properties of the solutions within each ensemble $\mathcal{E}^\mathrm{thin}_{n_\mathrm{b},n_\mathrm{c}}$ independently, yielding the orientation-resolved means
\begin{equation}
\mathring{\mathrm{E}}[\chi]_{n_\mathrm{b},n_\mathrm{c}} = \frac{\sum_{k\in  \mathcal{E}^\mathrm{thin}_{n_\mathrm{b},n_\mathrm{c}}} w_k\,\chi_k}{\sum_{k\in \mathcal{E}^\mathrm{thin}_{n_\mathrm{b},n_\mathrm{c}}} w_k} \, ,
\label{Eq_orientation_resolved_mean}
\end{equation}
with $\chi\equiv x,y,z, \mathit{D}_\mathrm{iso}, \mathit{D}_\Delta^2, \mathit{R}_2, \mathit{T}_2$. The short-hand notation "$\mathring{\mathrm{E}}[\chi]$" will be used for simplicity to describe the collection of orientation-resolved means $\mathring{\mathrm{E}}[\chi]_{n_\mathrm{b},n_\mathrm{c}}$ originating from all bootstrap solutions $n_\mathrm{b}$ and all clusters $n_\mathrm{c}$. As for Equation~\ref{Eq_ODF_metrics}, the orientation-resolved means of $T_2$ and $R_2$ in Equation~\ref{Eq_orientation_resolved_mean} are computed separately, as both quantities are commonly found in the MRI literature. The total weight associated with each orientation-resolved mean is given by
\begin{equation}
\mathring{w}_{n_\mathrm{b},n_\mathrm{c}} =  \sum_{k\in \mathcal{E}^\mathrm{thin}_{n_\mathrm{b},n_\mathrm{c}}} w_k\, .
\label{Eq_cluster_weight}
\end{equation}
Analogously to the way the radii of ODFs are thresholded during ODF-peak computations (see Section~\ref{Sec_ODF_peaks}), if a cluster $n_\mathrm{c}$ satisfies 
\begin{equation}
\frac{\sum_{n_\mathrm{b}}\mathring{w}_{n_\mathrm{b},n_\mathrm{c}}}{\sum_{n_\mathrm{b},n_\mathrm{c}}\mathring{w}_{n_\mathrm{b},n_\mathrm{c}}} \leq 0.1 \, ,
\label{Eq_MC_DPC_filtering}
\end{equation}
it is discarded as irrelevant and the clustering procedure is entirely repeated with $N_\mathrm{c} \mapsto N_\mathrm{c}-1$. To illustrate the resulting orientational clusters, Figure~\ref{Figure_MC-DPC}.E displays the cluster-specific collection of mean orientations, $\mathring{\mathrm{E}}[\mathbf{u}] \equiv (\mathring{\mathrm{E}}[x], \mathring{\mathrm{E}}[y], \mathring{\mathrm{E}}[z])$, for the three clusters estimated in an \textit{in-vivo} three-way crossing. A common change from Cartesian to spherical basis for $\mathring{\mathrm{E}}[\mathbf{u}]$ yields a collection of cluster-specific angular positions, denoted by $(\mathring{\mathrm{E}}[\theta], \mathring{\mathrm{E}}[\phi])$ for simplicity. Note that the intra-cluster averaging of Equation~\ref{Eq_orientation_resolved_mean} conceptually derives from the per-bin averaging of Refs.~\onlinecite{Topgaard:2019,deAlmeidaMartins:2020}, the main difference with these works being that the clusters are detected automatically instead of being defined by the user.

Finally, we extracted the median and interquartile range of our various orientation-resolved means across bootstrap solutions. To obtain the equivalent of a median cluster orientation, we computed the geometric median orientation $\mathbf{u}_{\mathrm{Med},n_\mathrm{c}}\equiv (\theta_{\mathrm{Med},n_\mathrm{c}}, \phi_{\mathrm{Med},n_\mathrm{c}})$ of each cluster-specific collection of mean orientations, $\{\mathring{\mathrm{E}}[\mathbf{u}]_{n_\mathrm{b},n_\mathrm{c}}\}_{1\leq n_\mathrm{b}\leq N_\mathrm{b}}$, \textit{via}
\begin{equation}
\mathbf{u}_{\mathrm{Med},n_\mathrm{c}}  = \underset{\mathbf{v}}{\mathrm{argmin}} \left[\sum_{n_\mathrm{b}=1}^{N_\mathrm{b}} \mathring{w}_{n_\mathrm{b},n_\mathrm{c}} \, d(\mathbf{v}, \mathring{\mathrm{E}}[\mathbf{u}]_{n_\mathrm{b},n_\mathrm{c}}) \right]
\label{Eq_geometric_median}
\end{equation}
using the angular distance $d$ defined in Equation~\ref{Eq_distance_MC_DPC}. Inspired by Ref.~\onlinecite{Jones:2003}, the equivalent of a cone of uncertainty around the geometric median orientation $\mathbf{u}_{\mathrm{Med},n_\mathrm{c}}$ was also computed, with half-aperture given by the median angular distance to $\mathbf{u}_{\mathrm{Med},n_\mathrm{c}}$ within a cluster:
\begin{equation}
\underset{n_\mathrm{b}}{\mathrm{Med}}\left(\mathring{\Delta}\beta_{n_\mathrm{b},n_\mathrm{c}}\right) = \underset{n_\mathrm{b}}{\mathrm{Med}}\left(  d(\mathbf{u}_{\mathrm{Med},n_\mathrm{c}}, \mathring{\mathrm{E}}[\mathbf{u}]_{n_\mathrm{b},n_\mathrm{c}}) \right) \, .
\label{Eq_cone_of_uncertainty}
\end{equation}
The practical realization of MC-DPC was achieved by extending the MATLAB toolbox of Refs.~\onlinecite{Nilsson_ISMRM:2018, Matlab_toolbox}.

\end{widetext}

\subsection{In silico evaluation of MC-DPC}
\label{Sec_in_silico_data}

An \textit{in silico} evaluation of MC-DPC was performed by designing a set of four components  $\{(\mathit{D}_{\parallel,\mathit{i}}, \mathit{D}_{\perp,\mathit{i}} , \theta_\mathit{i}, \phi_\mathit{i}, \mathit{R}_{2,\mathit{i}}, \mathit{w}_\mathit{i})\}_{1\leq i \leq N= 4}$ mimicking a three-way crossing and an isotropic component to simulate partial voluming:
\begin{itemize}
\item one isotropic component, $\mathit{D}_\mathrm{iso} =2$ {\textmu}m$^2$/ms, $\mathit{T}_2 = 1/\mathit{R}_2 = 500$ ms, $\mathit{w}_\mathrm{iso} = 0.1$.
\item one anisotropic component along $\mathit{x}$, $\mathit{D}_\mathrm{iso} = 0.9$ {\textmu}m$^2$/ms, $\mathit{D}_\Delta=\sqrt{0.75}=0.87$, $\mathit{R}_{2,x} = 1/\mathit{T}_{2,x}$, $\mathit{w} = (1-\mathit{w}_\mathrm{iso})/3$. 
\item one anisotropic component along $\mathit{y}$, $\mathit{D}_\mathrm{iso} = 0.8$ {\textmu}m$^2$/ms, $\mathit{D}_\Delta=\sqrt{0.8}=0.89$, $\mathit{R}_{2,y} = 1/\mathit{T}_{2,y}$, $\mathit{w} = (1-\mathit{w}_\mathrm{iso})/3$. 
\item one anisotropic component along $\mathit{z}$, $\mathit{D}_\mathrm{iso} = 0.7$ {\textmu}m$^2$/ms, $\mathit{D}_\Delta=\sqrt{0.85}=0.92$, $\mathit{R}_{2,z} = 1/\mathit{T}_{2,z}$, $\mathit{w} = (1-\mathit{w}_\mathrm{iso})/3$.
\end{itemize}
In order to assess MC-DPC's accuracy in estimating cluster-specific $T_2$ values, we replicated this set of four components into three configurations featuring the same diffusion properties, yet different $T_2$ values for the anisotropic components:
\begin{itemize}
\item configuration 1: $\mathit{T}_{2,x} = 70$ ms, $\mathit{T}_{2,y} = 100$ ms, $\mathit{T}_{2,z} = 90$ ms.
\item configuration 2: $\mathit{T}_{2,x} = 100$ ms, $\mathit{T}_{2,y} = 65$ ms, $\mathit{T}_{2,z} = 80$ ms.
\item configuration 3: $\mathit{T}_{2,x} = 60$ ms, $\mathit{T}_{2,y} = 75$ ms, $\mathit{T}_{2,z} = 90$ ms.
\end{itemize}

The ground-truth signals associated with these systems were computed using Equation~\ref{Eq_signal_discretized} and the acquisition scheme detailed in Section~\ref{Sec_in_vivo_data}. This calculation is in agreement with the conventional procedure for testing 1D, 2D or 4D Laplace inversion algorithms, where the ground-truth signal is calculated with the same kernel as the inversion.~\citep{Provencher:1982, Whittall:1989, Venkataramanan:2002, Mitchell:2012, Reymbaut_accuracy_precision:2020} Rician noise was then added to the ground-truth signals according to
\begin{equation}
\mathcal{S} = \sqrt{\left( \mathcal{S}_\mathrm{gt} + \frac{\nu}{\mathrm{SNR}} \right)^2 + \left(\frac{\nu^\prime}{\mathrm{SNR}} \right)^2} \, ,
\label{Eq_noisy_signals}
\end{equation}
where $\mathcal{S}_\mathrm{gt}$ is a ground-truth signal, $\mathcal{S}$ is the corresponding noisy signal, $\nu$ and $\nu^\prime$ denote random numbers drawn from a normal distribution with zero mean and unit standard deviation, and $\mathrm{SNR}$ is the desired signal-to-noise ratio. Three SNRs were considered: the $\mathrm{SNR}=90$ of our \textit{in vivo} dataset, a clinically relevant $\mathrm{SNR} = 30$, and an intermediate $\mathrm{SNR} = 70$. Since the Rician bias has been shown to be relevant only when the SNR is lower than five,~\citep{Gudbjartsson_Patz:1995} affecting the estimation of diffusion metrics,~\citep{Jones_Basser:2004,Gilbert:2007,Sotiropoulos:2013} the \textit{in silico} results of Section~\ref{Sec_in_silico_results} are identical to those yielded by signals with added Gaussian noise. For each of the aforementioned SNRs, the signals obtained from Equation~\ref{Eq_noisy_signals} were inverted using the 5D Monte-Carlo inversion detailed in Section~\ref{Sec_MC_inversion} with either $N_\mathrm{b}=50$, $N_\mathrm{b}=75$ or $N_\mathrm{b}=100$ bootstrap solutions. Indeed, varying $N_\mathrm{b}$ is relevant when validating MC-DPC because the number of data points that serve as its input is roughly proportional to $N_\mathrm{b}$, as seen from Equation~\ref{Eq_ensemble_thin_all_bootstrap}. Finally, MC-DPC was run on the solutions retrieved from each of these inversions and noise levels to extract orientation-resolved means.

\section{Results and discussion}
\label{Sec_Results}

\subsection{In silico evaluation}
\label{Sec_in_silico_results}

\begin{widetext}

\begin{figure*}[ht!]
\begin{center}
\includegraphics[width=42pc]{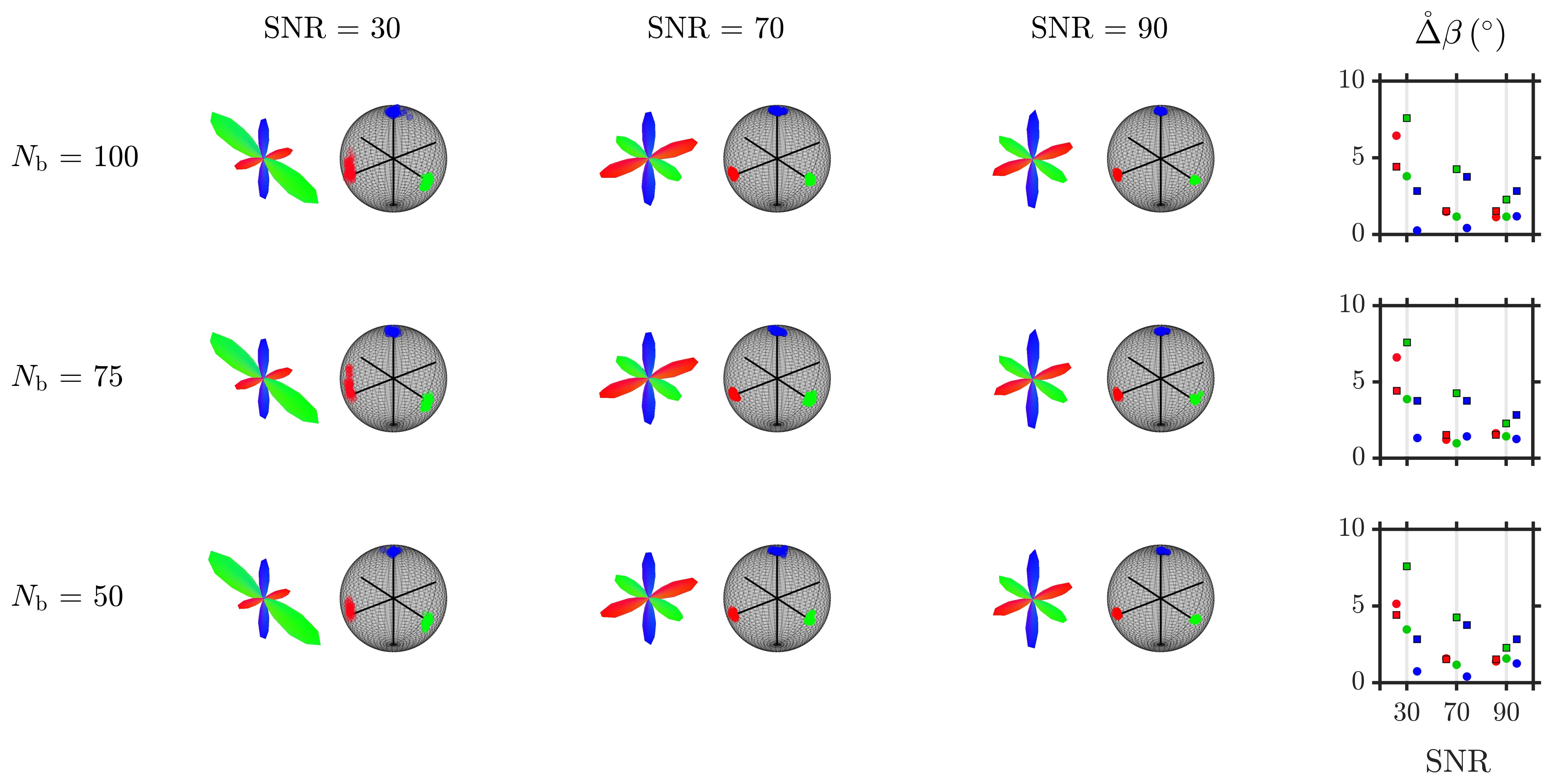}
\caption{Sub-voxel orientations retrieved for the \textit{in silico} data dubbed "configuration 1" in Section~\ref{Sec_in_silico_data} using the Monte-Carlo inversion for various numbers $N_\mathrm{b}$ of bootstrap solutions and various SNR levels. While the ODFs were obtained via the process detailed in Section~\ref{Sec_ODF_peaks}, the orientational clusters, here represented on the unit sphere, were extracted \textit{via} MC-DPC according to Section~\ref{Sec_MC-DPC}. $\mathring{\Delta}\beta$ denotes the angular deviation, computed for a given orientational cluster as the shortest angle between either the cluster geometric median orientation (circles, see Equation~\ref{Eq_geometric_median}) or the ODF peak (squares, see Section~\ref{Sec_ODF_peaks}), and the closest ground-truth anisotropic component orientation. Color/representation conventions are identical to those of Figures~\ref{Figure_ODF_pedagogic} and \ref{Figure_MC-DPC}. The conditions of the \textit{in vivo} study presented in Section~\ref{Sec_in_vivo_results} are closest to the case $(N_\mathrm{b}=100, \mathrm{SNR} =90)$.}
\label{Figure_in_silico_orientations}
\end{center}
\end{figure*}

\end{widetext}

\begin{widetext}

\begin{figure*}[ht!]
\begin{center}
\includegraphics[width=35pc]{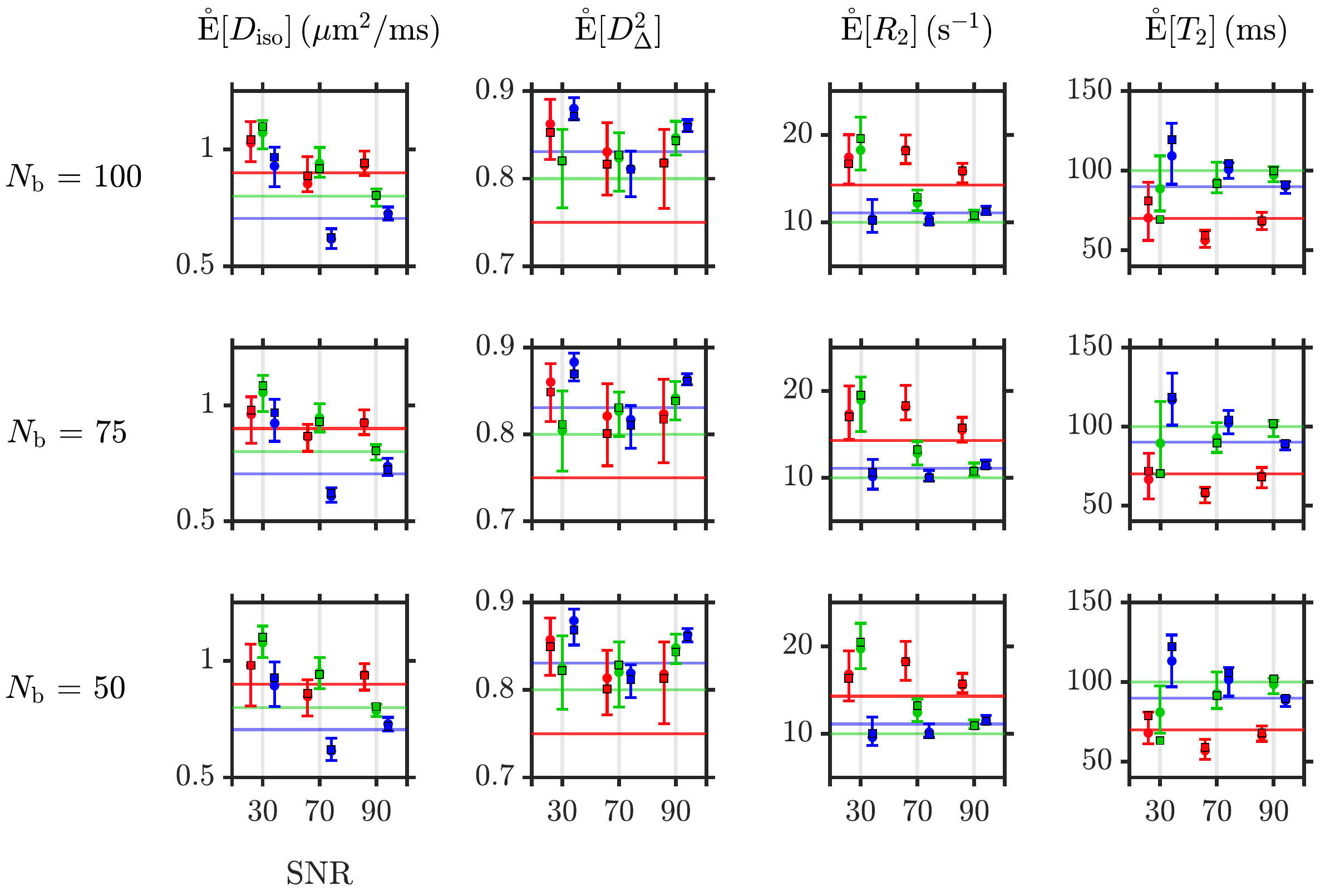}
\caption{Orientation-resolved means $\mathring{\mathrm{E}}[\chi]$ associated with the MC-DPC clusters of Figure~\ref{Figure_in_silico_orientations}. While ground-truth is shown as horizontal lines, the circles and whiskers represent the medians and interquartile ranges of the orientation-resolved means across bootstrap solutions, respectively. Squares correspond to the estimated ODF-peak metrics (see Equation~\ref{Eq_ODF_metrics}, evaluated at an ODF-peak orientation). Colors match those of the orientational clusters/ODF peaks presented in Figure~\ref{Figure_in_silico_orientations}. The conditions of the \textit{in vivo} study presented in Section~\ref{Sec_in_vivo_results} are closest to the case $(N_\mathrm{b}=100, \mathrm{SNR} =90)$.}
\label{Figure_in_silico_metrics}
\end{center}
\end{figure*}

\end{widetext}

\begin{figure*}[ht!]
\begin{center}
\includegraphics[width=42pc]{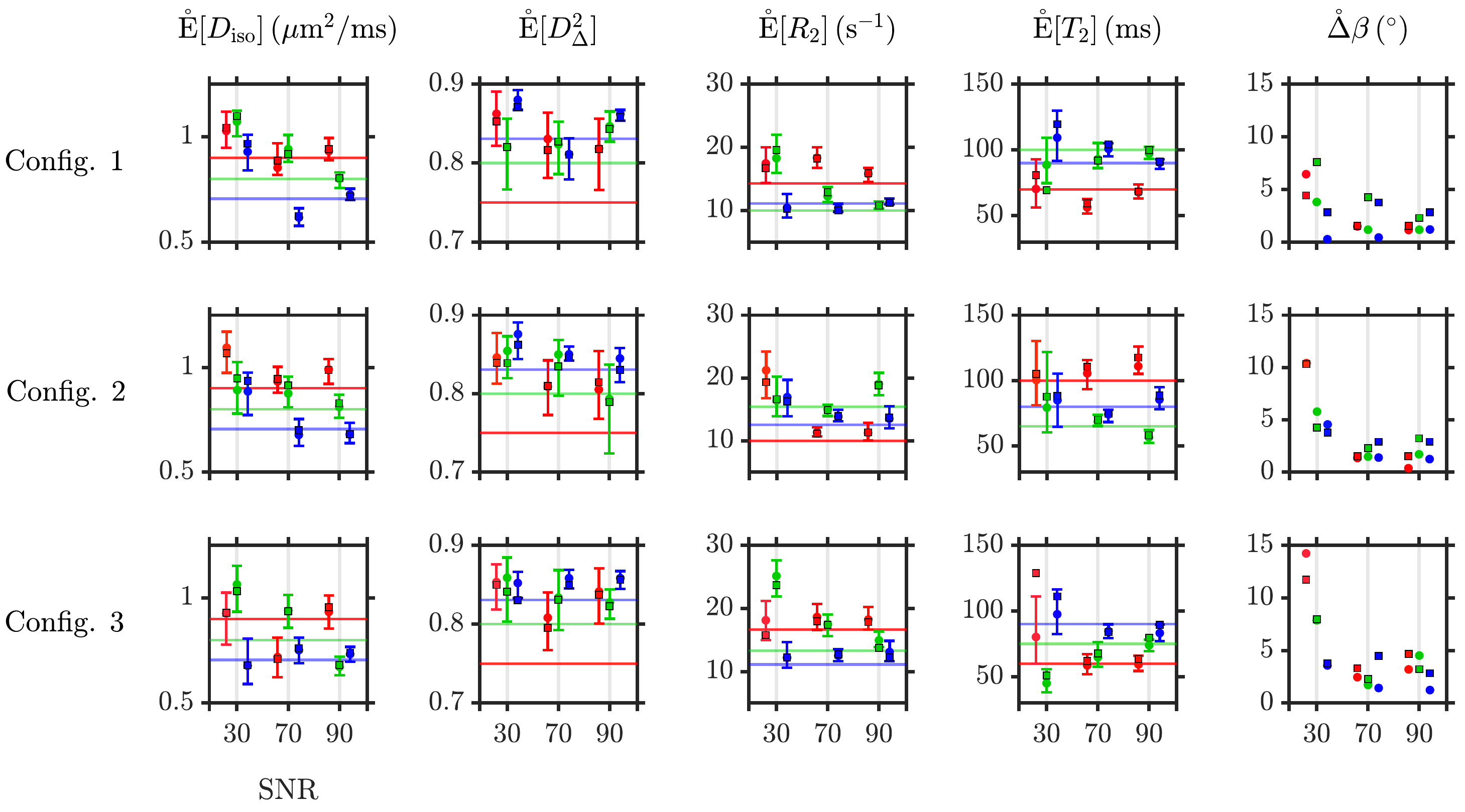}
\caption{Orientation-resolved means $\mathring{\mathrm{E}}[\chi]$ and angular deviation $\mathring{\Delta}\beta$ retrieved from MC-DPC at $N_\mathrm{b}=100$ for the three configurations presented in Section~\ref{Sec_in_silico_data}. Squares correspond to the estimated ODF-peak metrics (see Equation~\ref{Eq_ODF_metrics}, evaluated at an ODF-peak orientation). Color/representation conventions are identical to those of Figures~\ref{Figure_in_silico_orientations} and \ref{Figure_in_silico_metrics}. The conditions of the \textit{in vivo} study presented in Section~\ref{Sec_in_vivo_results} are closest to the $\mathrm{SNR} =90$ level.}
\label{Figure_in_silico_metrics_configs}
\end{center}
\end{figure*}

Figure~\ref{Figure_in_silico_orientations} displays the sub-voxel orientations retrieved for the \textit{in silico} data dubbed "configuration 1" in Section~\ref{Sec_in_silico_data} using both the ODFs of Section~\ref{Sec_ODF_peaks} and MC-DPC from Section~\ref{Sec_MC-DPC}, and quantifies the angular deviation $\mathring{\Delta}\beta$ of each cluster geometric median orientation (see Equation~\ref{Eq_geometric_median}) compared to the closest ground-truth anisotropic component orientation. This angular deviation is not to be confused with the orientational uncertainty defined in Equation~\ref{Eq_cone_of_uncertainty}, as $\mathring{\Delta}\beta$ reflects an orientational bias. Figure~\ref{Figure_in_silico_metrics} quantifies the orientation-resolved means $\mathring{\mathrm{E}}[\chi]$ for the same \textit{in silico} data. Figure~\ref{Figure_in_silico_metrics_configs} quantifies the orientation-resolved means $\mathring{\mathrm{E}}[\chi]$ and angular deviation $\mathring{\Delta}\beta$ associated with the clusters estimated by MC-DPC for all three \textit{in silico} configurations detailed in Section~\ref{Sec_in_silico_data}, enabling to assess MC-DPC's accuracy in estimating cluster-specific $T_2$ values. ODF-peak information is also displayed on Figures~\ref{Figure_in_silico_orientations}, \ref{Figure_in_silico_metrics} and \ref{Figure_in_silico_metrics_configs} for comparison.

Before discussing Figures~\ref{Figure_in_silico_orientations}, \ref{Figure_in_silico_metrics} and \ref{Figure_in_silico_metrics_configs}, it is important to mention that the \textit{in silico} results have been generated for a single noise realization at their respective SNRs. When multiple noise realizations are considered at a fixed SNR, it appears that MC-DPC yields consistent estimations at $\mathrm{SNR}=90$, but not at $\mathrm{SNR}=30$ and $\mathrm{SNR}=70$, as shown in Appendix~\ref{App_in_silico_noise_realizations}. Given that MC-DPC merely takes the output of the Monte-Carlo inversion as input, this lack of consistency at low SNR originates from the 5D Monte-Carlo inversion itself, in the context of the acquisition scheme described in Section~\ref{Sec_in_vivo_data}. At lower dimension, the accuracy and precision of the 4D Monte-Carlo inversion have already been shown \textit{in silico} to be noise-sensitive.~\citep{Reymbaut_accuracy_precision:2020} For that reason, we mostly discuss Figures~\ref{Figure_in_silico_orientations}, \ref{Figure_in_silico_metrics} and \ref{Figure_in_silico_metrics_configs} at $\mathrm{SNR}=90$ below.

Within Figure~\ref{Figure_in_silico_orientations}, the orientational spread of the MC-DPC clusters informs on the precision of the underlying Monte-Carlo inversion. Clusters become more orientationally dispersed as $\mathrm{SNR}$ decreases at constant $N_\mathrm{b}$, which is associated with the loss of precision of the 5D Monte-Carlo inversion with reduced $\mathrm{SNR}$. At constant SNR, the orientational clusters retrieved from MC-DPC are rather unaffected upon changing $N_\mathrm{b}$. As for the angular deviation $\mathring{\Delta}\beta$, informing on the accuracy of the underlying Monte-Carlo inversion, it roughly reflects the same aforementioned trends on precision. Note that the $\mathrm{SNR}=30$ ODFs clearly violate the equiprobability of the ground-truth anisotropic components of Section~\ref{Sec_in_silico_data}, indicating a poor performance of the Monte-Carlo inversion (see Appendix~\ref{App_in_silico_noise_realizations}). Compared to ODF-peak orientations, cluster geometric median orientations seem to be more accurate as they typically yield smaller values of $\mathring{\Delta}\beta$, probably due to the fact that our ODFs are bound to a given discrete spherical mesh (see Section~\ref{Sec_ODF_peaks}).

Regarding Figure~\ref{Figure_in_silico_metrics}, the estimations of the orientation-resolved means suffer from the same loss of accuracy and precision mentioned for the orientational information in Figure~\ref{Figure_in_silico_orientations} as $\mathrm{SNR}$ decreases at constant $N_\mathrm{b}$, except for $\mathring{\mathrm{E}}[D_\Delta^2]$'s estimations, whose precision is unaffected across $\mathrm{SNR}$ levels. Nevertheless, the $(N_\mathrm{b}=100, \mathrm{SNR} =90)$ case, \textit{i.e.} that closest to the following \textit{in vivo} study, appears to yield rather accurate estimations of the changes in $\mathring{\mathrm{E}}[R_2]$ and $\mathring{\mathrm{E}}[T_2]$ across fiber populations while yielding satisfying orders of magnitude for $\mathring{\mathrm{E}}[D_\mathrm{iso}]$ and $\mathring{\mathrm{E}}[D_\Delta^2]$. At such high $\mathrm{SNR}$, the number of bootstrap solutions $N_\mathrm{b}$ seems to have mild to negligible effect on the MC-DPC results. As for ODF-peak metrics, they tend to agree with the median orientation-resolved means at intermediate and high SNR levels, but may shift away from these medians at low SNR.

Finally, Figure~\ref{Figure_in_silico_metrics_configs} demonstrates that estimated orientation-resolved mean $T_2$ values agree well with the ground truth across \textit{in silico} configurations at $\mathrm{SNR}=90$, thereby establishing MC-DPC's accuracy in estimating cluster-specific transverse relaxation properties at high SNR. For that SNR level, $\mathring{\Delta}\beta$ remains below five degrees for all estimated clusters. As for ODF-peak metrics and orientations, they present the same features as those discussed in Figures~\ref{Figure_in_silico_orientations} and \ref{Figure_in_silico_metrics}.

\subsection{In vivo evaluation}
\label{Sec_in_vivo_results}

\begin{figure*}[ht!]
\begin{center}
\includegraphics[width=34pc]{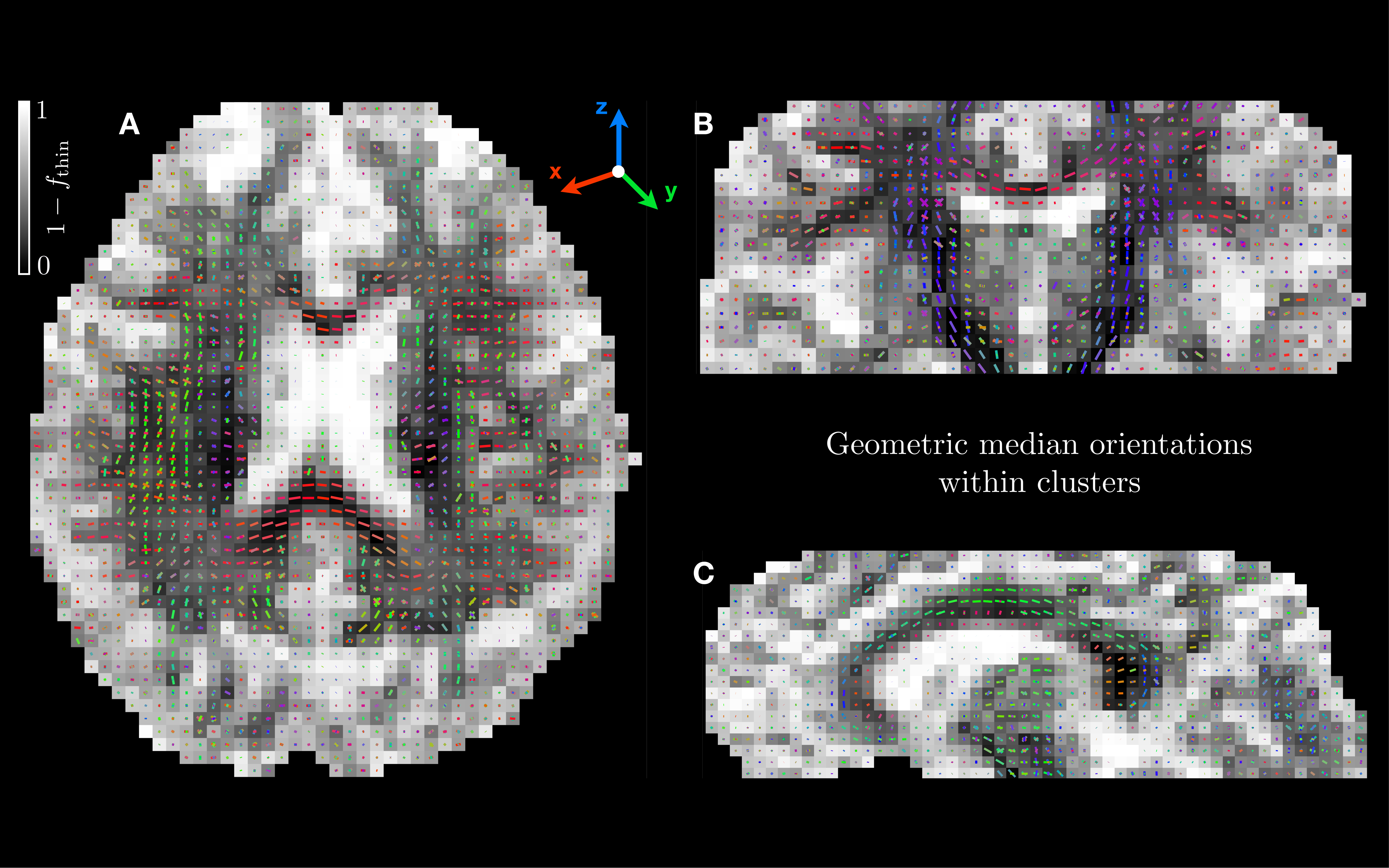}
\caption{Peak map associated with the geometric median orientations (see Equation~\ref{Eq_geometric_median}) of the clusters retrieved with MC-DPC in typical axial (A), coronal (B) and sagittal (C) slices. Peak color codes for orientation, with $x$, $y$ and $z$ corresponding to the "left-right", "anterior-posterior" and "inferior-superior" directions, respectively. For a given cluster $n_\mathrm{c}$, the peak norm is weighted by the median cluster weight $\mathrm{Med}_{n_\mathrm{b}}(\mathring{w}_{n_\mathrm{b},n_\mathrm{c}})$ (see Equation~\ref{Eq_cluster_weight}) and by the voxel-scale fraction of thin-bin solutions. Denoting by $f_\mathrm{thin}$ the total signal fraction of solutions that belong to the thin bin, the greyscale map displays the total signal fraction of solutions that do not belong to the thin bin.}
\label{Figure_local_orientations}
\end{center}
\end{figure*}

\begin{figure*}[ht!]
\begin{center}
\includegraphics[width=34pc]{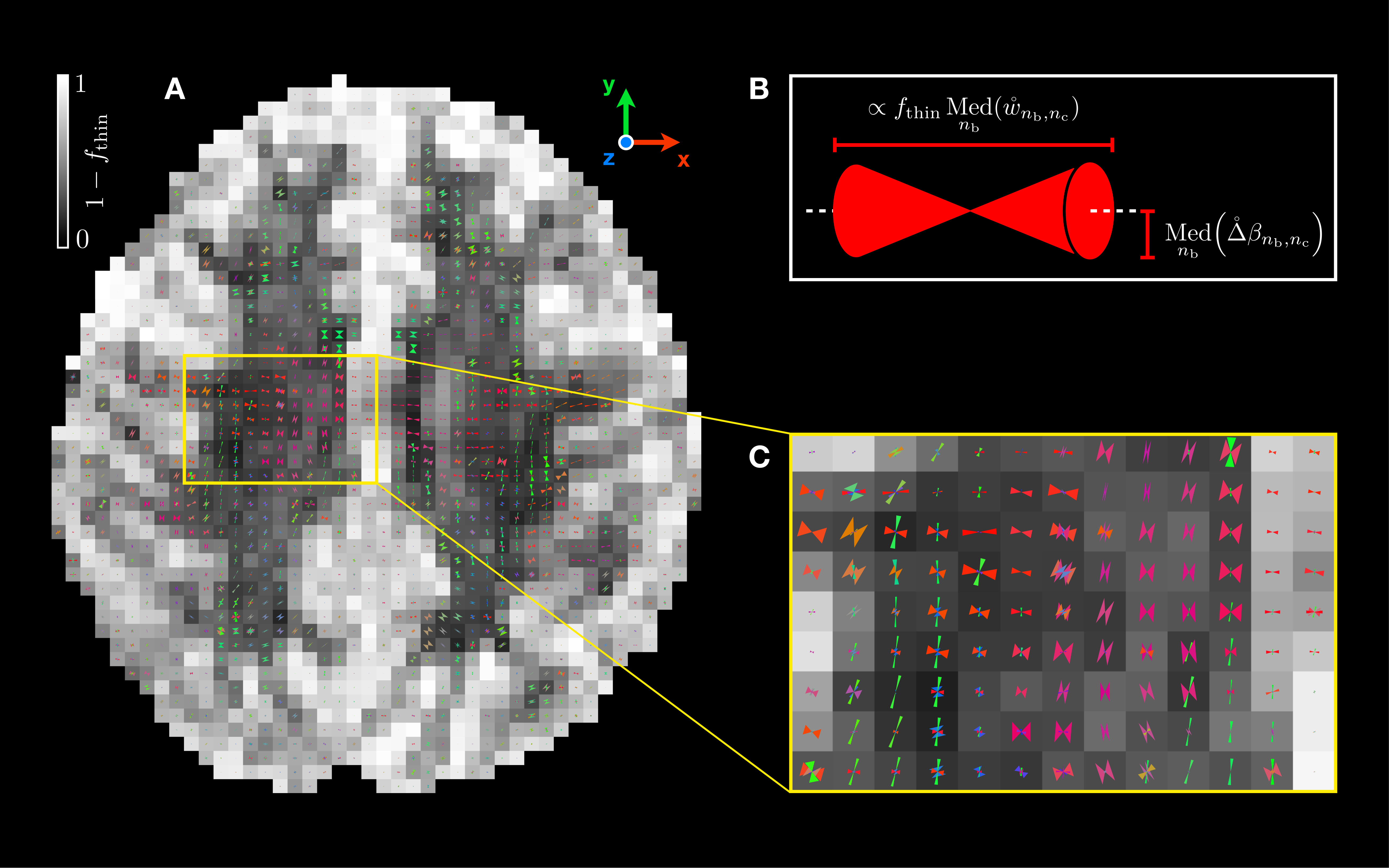}
\caption{(A) Cones of uncertainty oriented along the geometric median orientations (see Equation~\ref{Eq_geometric_median}) of the clusters retrieved with MC-DPC in a typical axial slice. The greyscale map displays the total signal fraction of solutions that do not belong to the thin bin. (B) Representation convention for the cones of uncertainty, displayed as bi-triangular projections in panels (A) and (C). While the length of a given cone is weighted by the median cluster weight $\mathrm{Med}_{n_\mathrm{b}}(\mathring{w}_{n_\mathrm{b},n_\mathrm{c}})$ (see Equation~\ref{Eq_cluster_weight}) and by the voxel-scale fraction $f_\mathrm{thin}$ of thin-bin solutions, its half-aperture corresponds to the median angular distance to $(\theta_{\mathrm{Med},n_\mathrm{c}}, \phi_{\mathrm{Med},n_\mathrm{c}})$ within the associated cluster, $\mathrm{Med}_{n_\mathrm{b}}(\mathring{\Delta}\beta_{n_\mathrm{b},n_\mathrm{c}})$ (see Equation~\ref{Eq_cone_of_uncertainty}). (C) Zoom on an area of crossings in the corona radiata. This area primarily features crossings between the corpus callosum, arcuate fasciculus and corticospinal tract, and between the corpus callosum and cingulum. The cones of uncertainty, distinct for each cluster, appear to capture white-matter fanning \textit{via} an increased aperture, as seen in the fanning part of the corpus callosum.}
\label{Figure_local_cones}
\end{center}
\end{figure*}

Figure~\ref{Figure_local_orientations} presents peak maps corresponding to the geometric median orientations (see Equation~\ref{Eq_geometric_median}) of the clusters retrieved with MC-DPC in the \textit{in vivo} dataset described in Section~\ref{Sec_in_vivo_data}. The cluster local orientations are consistent with the known white-matter anatomy, highlighting main fiber bundles such as the corticospinal tract, corpus callosum, arcuate fasciculus and cingulum, even in areas of crossings such as the corona radiata. Figure~\ref{Figure_local_cones} extends the orientational information of Figure~\ref{Figure_local_orientations} by showing the cones of uncertainty estimated by MC-DPC, as defined in Equation~\ref{Eq_cone_of_uncertainty}, particularly focusing on the corona radiata. These cones, distinct for each cluster, appear to capture white-matter fanning \textit{via} an increased aperture (\textit{c.f.} corpus callosum).

\begin{figure*}[ht!]
\begin{center}
\includegraphics[width=35pc]{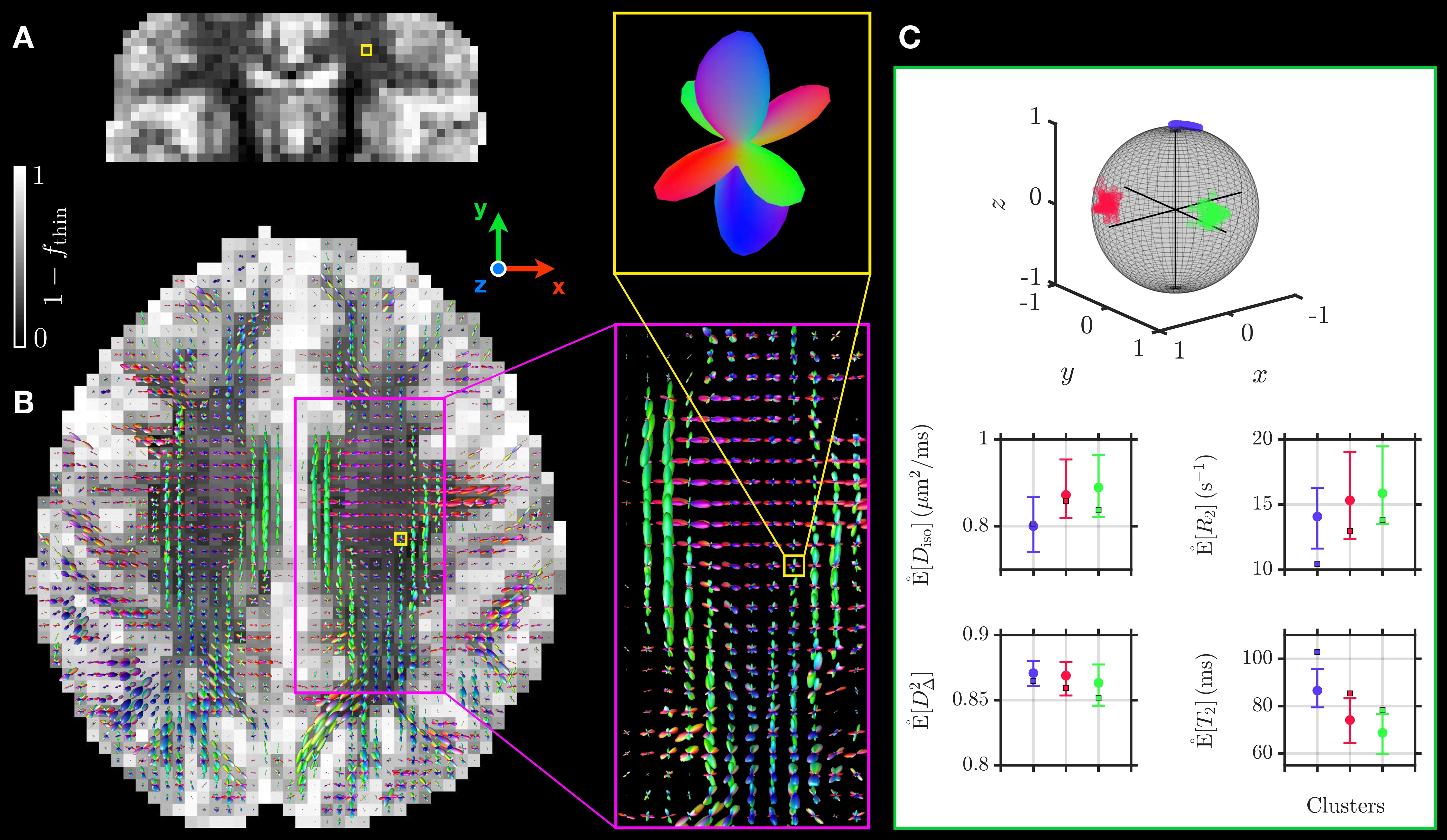}
\caption{Orientation-resolved means in a typical corona-radiata voxel, chosen from the area of crossing between the corpus callosum, the arcuate fasciculus and the corticospinal tract. (A) Coronal slice containing the voxel of interest (yellow square) and showing the signal fraction map of non thin-bin components (see Figure~\ref{Figure_local_orientations}). (B) Axial slice containing the same voxel and showing the same map, with superimposed ODFs. Color codes for orientation, with $x$, $y$ and $z$ corresponding to the "left-right", "anterior-posterior" and "inferior-superior" directions, respectively. Pink and yellow insets zoom on an area of interest and on the ODF in the voxel of interest, respectively. (C) Orientation-resolved means in the voxel of interest. The red, green and blue clusters correspond to the corpus callosum, arcuate fasciculus and corticospinal tract, respectively. Square points indicate the ODF-peak metrics shown in Figure~\ref{Figure_ODF_pedagogic}. Color/representation conventions are identical to those of Figures~\ref{Figure_MC-DPC}, \ref{Figure_in_silico_orientations} and \ref{Figure_in_silico_metrics}.}
\label{Figure_CR}
\end{center}
\end{figure*}

Figure~\ref{Figure_CR} displays the orientation-resolved means $\mathring{\mathrm{E}}[\chi]$ estimated in a typical voxel in the corona radiata (three-way fiber crossing). These means can directly be compared to the ODF-peak metrics of Figure~\ref{Figure_ODF_pedagogic}. While the ODF-peak metrics qualitatively convey similar differences between fiber populations, they do not agree quantitatively with the median orientation-resolved means of MC-DPC, which was not the case \textit{in silico} for the high SNR characteristic of our \textit{in vivo} dataset (see Figures~\ref{Figure_in_silico_metrics} and \ref{Figure_in_silico_metrics_configs}). This mismatch may come from more realistic features of the voxel content, \textit{e.g.} fiber dispersion or intra-fiber heterogeneity, that are hard to account for \textit{in silico} and may in turn further exaggerate the difference in performance observed \textit{in silico} between the two methods at lower SNR levels.

Besides, unlike ODF-peak metrics (see Equation~\ref{Eq_ODF_metrics}, evaluated at an ODF-peak orientation), orientation-resolved means offer the possibility to perform statistical tests such as non-parametric Mann-Whitney $U$ tests and non-parametric Kruskal-Wallis tests to potentially reject null hypotheses regarding identical medians and identical distributions for $\mathring{\mathrm{E}}[\chi]$ across clusters, respectively. For each type of orientation-resolved mean presented in Figure~\ref{Figure_CR}, such tests reject the aforementioned null hypotheses at the 1\% significance level between any pair of clusters, except for $\mathring{\mathrm{E}}[D_\Delta^2]$ between the corpus-callosum and corticospinal-tract clusters, which yields a $p$-value of $0.029$ for both the Mann-Whitney $U$ test and the two-group Kruskal-Wallis test. 


\begin{figure*}[ht!]
\begin{center}
\includegraphics[width=35pc]{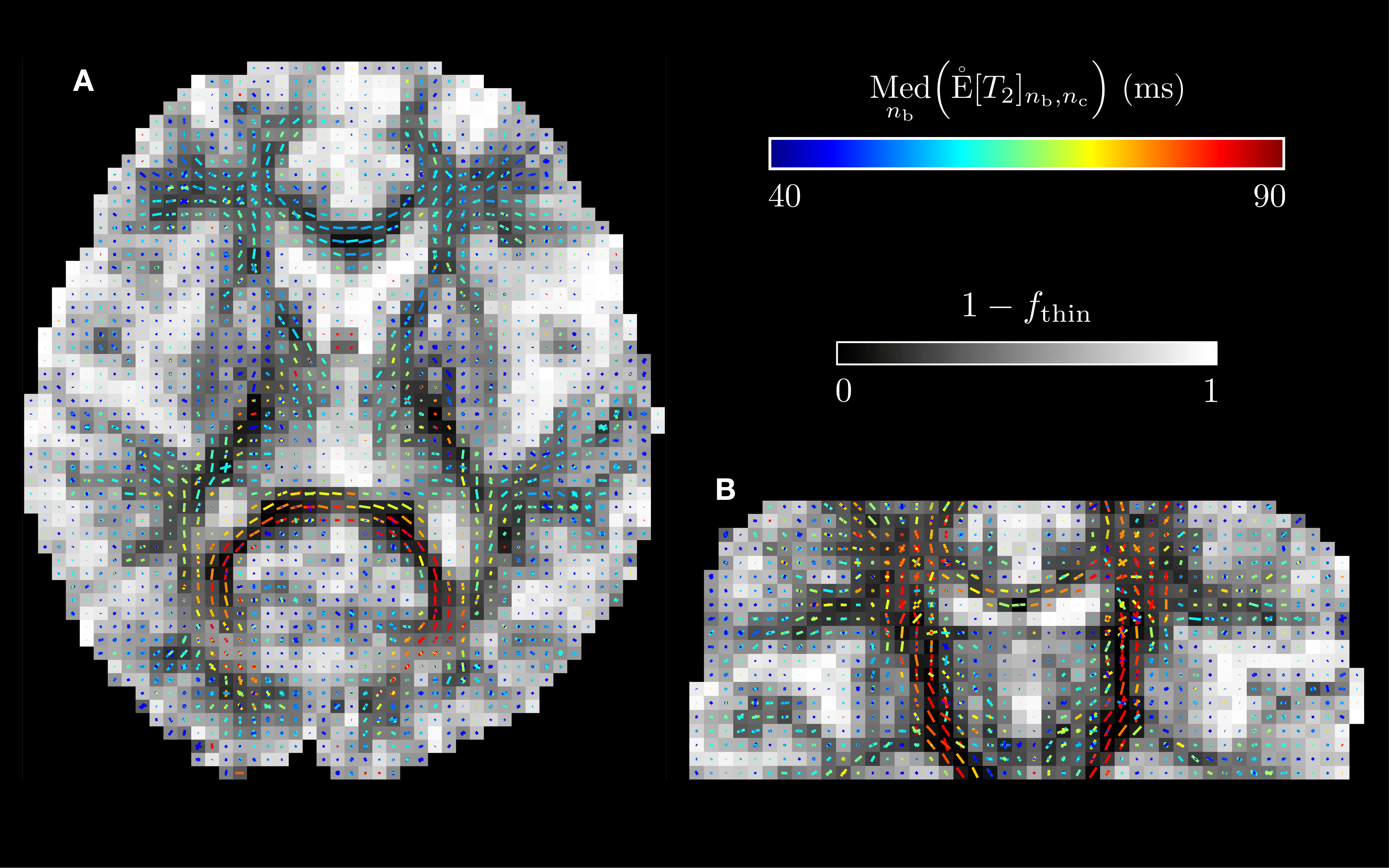}
\caption{Peak map associated with the geometric median orientations (see Equation~\ref{Eq_geometric_median}) of the clusters retrieved with MC-DPC in an axial slice containing the forceps major (A), and a coronal slice containing the corticospinal tract (B). Peak color codes for the median orientation-resolved $T_2$ within each cluster. Representation conventions are identical to those of Figure~\ref{Figure_local_orientations}. The spatial distribution of the red-colored peaks suggests that the median orientation-resolved $T_2$ is consistently largest along the corticospinal tract and the forceps major.}
\label{Figure_peak_T2}
\end{center}
\end{figure*}

In particular, the rejection of the null hypotheses for $\mathring{\mathrm{E}}[\mathit{T}_2]$ hints at a significantly higher $\mathring{\mathrm{E}}[\mathit{T}_2]$ in the corticospinal-tract cluster than in the corpus-callosum and arcuate-fasciculus clusters. This observation is pushed beyond the single-voxel level in Figure~\ref{Figure_peak_T2}, which shows peak maps similar to those of Figure~\ref{Figure_local_orientations}, but colored with the median orientation-resolved $T_2$ within each cluster. This figure indicates that $T_2$ is consistently largest along the corticospinal tract and the forceps major. A similar observation was made for the corticospinal tract in Ref.~\onlinecite{Lampinen:2020}. From a biological perspective, these bundles appear to differ from other bundles in two ways that may explain these longer transverse relaxation times. Firstly, their high fractions of large axons \citep{dellAcqua_ISMRM:2019} could result in differential surface relaxation effects. Second, their high myelin-water fractions \citep{Liu:2019} could lead to increased $T_2$ values \textit{via} slower exchange between the axon-water and myelin-water pools.~\citep{Dula:2010}

\vspace*{\fill}

\newpage 
\clearpage
\newpage

\section{Conclusions}

The Monte-Carlo density-peak clustering (MC-DPC) procedure builds upon the non-parametric distributions obtained by Monte-Carlo inversions and extracts orientation-resolved means quantifying the median value and interquartile range of diffusion-relaxation properties within orientational clusters associated with sub-voxel fiber orientations. While our \textit{in silico} evaluation of MC-DPC features accurate estimations at high SNR, our \textit{in vivo} evaluation on diffusion- and $T_2$-weighted datasets demonstrates that it captures multiple sub-voxel fiber orientations, fiber-specific cones of uncertainty, and fiber-specific diffusion-relaxation properties consistent with the known anatomy and existing literature. 

As such, MC-DPC shows potential in multiple avenues of exploring and quantifying fiber-specific microstructure. Firstly, it could enable monitoring either group differences in tissue microstructure within a sub-voxel fiber population, or fiber-specific microstructural changes in longitudinal studies, such as in development, aging or treatment response. Secondly, it could increase specificity to particular white-matter fiber bundles, even in areas of partial voluming (given the assumption-free aspect of the Monte-Carlo inversion), which is pertinent to tease apart crossing fibers in tractography \citep{deSantis_T1:2016, Andrews_ISMRM:2019, Lampinen:2020} and to track around brain tumors during surgical planning. Thirdly, the output of MC-DPC could be adapted to serve as input for the convex optimization modeling for microstructure-informed tractography (COMMIT) framework,~\citep{Daducci_COMMIT:2015,Daducci:2016,Schiavi:2019,Barakovic_thesis:2019} which filters out tracks that do not satisfy some input prior information. Additional applications of MC-DPC would lie in its straightforward translation from diffusion-$T_2$ datasets to diffusion-$T_1$ and diffusion-$T_1$-$T_2$ datasets, which are relevant to tract-specific myelination mapping,~\citep{deSantis_T1:2016, Andrews_ISMRM:2019} and to study the white-matter angular dependence of $\mathit{T}_1$ \citep{Henkelman:1994,Knight:2018} and $\mathit{T}_2^{(*)}$ \citep{Henkelman:1994, Bender_Klose:2010, Lee:2011, Rudko:2014, Knight:2015, Gil:2016, McKinnon:2019} with respect to the main MRI magnetic field $\mathbf{B}_0$. 

As a more global outlook, MC-DPC's reduced performance at low to intermediate $\mathrm{SNR}$, reflecting that of the 5D Monte-Carlo inversion itself at such SNR levels, highlights the crucial importance of optimizing acquisition protocols to yield better non-parametric signal inversions. Even though multiple works have established protocol-optimization procedures targeting specific parametric models,~\citep{Cercignani_Alexander:2006, Alexander:2008, Alexander:2010, Lampinen:2017, Coelho:2019, Coelho:2019_MICCAI, Lampinen:2020} much effort has yet to be spent in establishing similar procedures to target non-parametric signal inversions.~\citep{Song:2005,Bates_ISMRM:2019,Song:2020}


\begin{appendices}


\section{Describing the voxel content by a distribution of diffusion-relaxation features}
\label{App_DTD_description}

The measured signal $\mathcal{S}$ associated with a diffusion-relaxation correlation experiment probes diffusion-relaxation processes over specific observational time-scales that depend on the choice of experimental time parameters. For given observational time-scales, a common description of the sub-voxel composition of heterogeneous tissues is obtained by considering a "snapshot" of the combined non-Gaussian diffusion effects of restriction and exchange, and by approximating the signal decay as a continuous weighted sum of exponential decays, giving the following multidimensional Laplace transform:~\citep{deAlmeidaMartins_Topgaard:2018}
\begin{widetext}
\begin{equation}
\frac{\mathcal{S}(\mathbf{b},\bm{\tau})}{\mathcal{S}_0} = \int_0^{+\infty}\int_0^{+\infty}\int_{\mathrm{Sym}^{+}(3)} \! \mathcal{P}(\mathbf{D},\mathit{R}_2,\mathit{R}_1)\, \mathcal{K}(\mathbf{D},\mathit{R}_2,\mathit{R}_1, \mathbf{b},\bm{\tau}) \, \mathrm{d}\mathbf{D}\, \mathrm{d}\mathit{R}_2\, \mathrm{d}\mathit{R}_1 \, ,
\label{Eq_signal_tensor_distribution}
\end{equation}
\end{widetext}
where $\mathbf{b}$ is the diffusion-encoding tensor from tensor-valued diffusion encoding,~\citep{Eriksson:2013,Westin:2014,Eriksson:2015,Westin:2016,Topgaard:2017,Topgaard_dim_rand_walks:2019} $\bm{\tau}$ represents the experimental time parameters linked to relaxation, $\mathcal{P}(\mathbf{D},\mathit{R}_2,\mathit{R}_1)$ is the joint distribution of apparent diffusion tensors~\citep{Jian:2007} $\mathbf{D}$ and apparent relaxation rates $\mathit{R}_2 = 1/\mathit{T}_2$ and $\mathit{R}_1 = 1/\mathit{T}_1$, $\mathcal{K}(\mathbf{D},\mathit{R}_2,\mathit{R}_1, \mathbf{b},\bm{\tau})$ is the integral kernel yielding the signal decay associated with a given set of experimental parameters $(\mathbf{b},\bm{\tau})$ and a given microscopic domain $(\mathbf{D},\mathit{R}_2,\mathit{R}_1)$, and $\mathcal{S}_0$ is the signal acquired for $(\mathbf{b},\bm{\tau})$ such that $\mathcal{K}(\mathbf{D},\mathit{R}_2,\mathit{R}_1, \mathbf{b},\bm{\tau}) = 1$ for all $(\mathbf{D},\mathit{R}_2,\mathit{R}_1)$. Here, $\mathrm{Sym}^{+}(3)$ denotes the space of symmetric positive-semidefinite 3$\times$3 tensors.

Changing the observational time-scales may very well lead to a different set of exponential decays as a result of restricted diffusion \citep{Woessner:1963} and exchange,~\citep{Johnson:1993,Li_Springer:2019} which implies that the measured $\mathcal{P}(\mathbf{D},\mathit{R}_2,\mathit{R}_1)$ may depend on the spectral content of the diffusion-encoding gradients.~\citep{Stepisnik:1981, Stepisnik:1985, Callaghan_Stepisnik:1995} Even though such time-dependent effects have been measured in human-brain white matter,~\citep{Van:2014,Baron_Beaulieu:2014,Baron_Beaulieu:2015,Fieremans:2016,Veraart:2019,Lundell:2019,dellAcqua_ISMRM:2019} spinal cord \citep{Jespersen:2018,Grussu:2019} and prostate \citep{Lemberskiy:2017,Lemberskiy:2018} using pulse sequences specifically designed for varying the observational time-scales over extended ranges, the above $\mathcal{P}(\mathbf{D},\mathit{R}_2,\mathit{R}_1)$ description holds for the limited range of long diffusion times probed by clinical dMRI experiments in the brain.~\citep{Clark:2001,Ronen:2006,Nilsson:2009, Nilsson:2013a, Nilsson:2013b, deSantis_T1:2016, Lampinen:2017,Veraart:2018,Grussu:2019,Szczepankiewicz_ISMRM:2019}

\section{Setting the concentration parameter of the ODF Watson kernel}
\label{App_kappa_ODF}

In this work, we considered a $1000$-point uniform spherical mesh within which the median minimal angular distance between two nearest-neighbor points roughly equals $7^\circ$. Denoting by $\beta_{\bm{\mu},\mathbf{u}_i}$ the shortest angle between $\bm{\mu}(\theta_\text{mesh},\phi_\text{mesh})$ and $\mathbf{u}_i$ in Equation~\ref{Eq_per_bootstrap_ODF}, one can use the small-angle asymptotic equivalent $\cos^2\beta \sim 1-\beta^2$ to write for small angles $\beta_{\bm{\mu},\mathbf{u}_i}$
\begin{widetext}
\begin{equation}
\exp (\kappa\, [\mathbf{u}_\mathit{i}\cdot\bm{\mu}(\theta_\text{mesh},\phi_\text{mesh})]^2) = \exp (\kappa\, \cos^2\beta_{\bm{\mu},\mathbf{u}_i}) \sim \exp(\kappa)\, \exp (-\kappa\beta_{\bm{\mu},\mathbf{u}_i}^2)\, .
\label{Eq_Watson_to_gaussian}
\end{equation}
\end{widetext}
This rewriting of the Watson kernel allows the value of $\kappa$ to be set intuitively so that the Watson kernel does not induce any significant peak broadening larger than the distance between two nearest-neighboring mesh nodes, as $\sigma= 1/\sqrt{2\kappa}$ yields the standard deviation of the Gaussian kernel in Equation~\ref{Eq_Watson_to_gaussian}. Here, we chose to set $\sigma$ to one and a half median minimal angular distance between two nearest-neighbor mesh points, \textit{i.e.} $10.5^\circ$, to prevent over-smoothing, which gives $\kappa=14.9$. This value of $\kappa$ corresponds to an orientation dispersion index~\citep{Zhang_NODDI:2012} $\mathrm{OD} = (2/\pi)\arctan(1/\kappa) = 0.04$ and an orientation order parameter~\citep{Lasic:2014,Topgaard_liquid:2016} $\mathrm{OP} = [\mathcal{M}(3/2,5/2,\kappa)/\mathcal{M}(1/2,3/2,\kappa) - 1]/2 = 0.89$, where $\mathcal{M}$ is Kummer's confluent hypergeometric function. 

\section{Brief review of clustering techniques}
\label{App_clustering_techniques}

The $K$-means \citep{Macqueen:1967} and $K$-medoids \citep{Kaufman:1987} methods both attempt to iteratively find the best cluster centroid (cluster center) by minimizing the distance between points labeled to be in a cluster and a point designated as the cluster centroid.~\citep{Kaufman:1987,Hoppner:1999,Frey:2007} However, these methods cannot detect non-spherical clusters nor outliers.~\citep{Jain:2010}
Another popular approach, mixture model-based clustering,~\citep{McLachlan:1988, McLachlan:2004, McLachlan:2005, McLachlan:2007} assumes that the data are sampled from a finite mixture of underlying probability distributions, with each distribution corresponding to a different cluster, but the accuracy of this approach depends on the ability of the trial probability distribution to represent the data.~\citep{Fraley:2002}
Hierarchical clustering,~\citep{Ward:1963} in contrast to the above partition-based clustering techniques, generates a hierarchical series of nested clusters. The output of this clustering class can be represented as a cluster tree called a "dendrogram", \textit{i.e.} a multi-level hierarchy of clusters within which clusters at one level are formed of smaller clusters at the next level. A specified number of clusters is obtained by cutting the dendrogram at a certain level. However, a major drawback of hierarchical clustering is its lack of robustness to noise in the data.~\citep{Zhang:2004}

Unlike other approaches, clusters of arbitrary shape and outliers can more straightforwardly be captured by approaches based on the local density of data points, such as the mean-shift clustering method,~\citep{Fukunaga:1975,Cheng:1995} the Density-based spatial clustering of applications with noise (DBSCAN) algorithm,~\citep{Ester:1996} the Ordering points to identify the clustering structure (OPTICS) algorithm,~\citep{Ankerst:1999} and the Density-based clustering (DENCLUE) algorithm.~\citep{Campello:2015} In DBSCAN, a fixed density threshold is set to discard as outliers the points whose density is lower than this threshold and to assign the other points to disconnected regions of high density. However, the choice of arbitrary threshold is non-trivial, a drawback that is not shared by the mean-shift clustering, a mode-seeking algorithm that iteratively locates local maxima of a density function defined on the space of data points. Yet, the mean-shift clustering algorithm can be computationally costly. Finally, a faster density-based clustering approach, called "density-peak clustering" (DPC),~\citep{Rodriguez_Laio:2014} combines the advantages of the $K$-medoids, DBSCAN and mean-shift methods while keeping the drawbacks of density-based approaches, \textit{i.e.} it relies solely on a distance between data points and can detect non-spherical clusters and outliers, but also requires arbitrary parameter choices. This is why we chose to base our work on DPC.

\section{In silico results across noise realizations}
\label{App_in_silico_noise_realizations}

\begin{figure*}[ht!]
\begin{center}
\includegraphics[width=37pc]{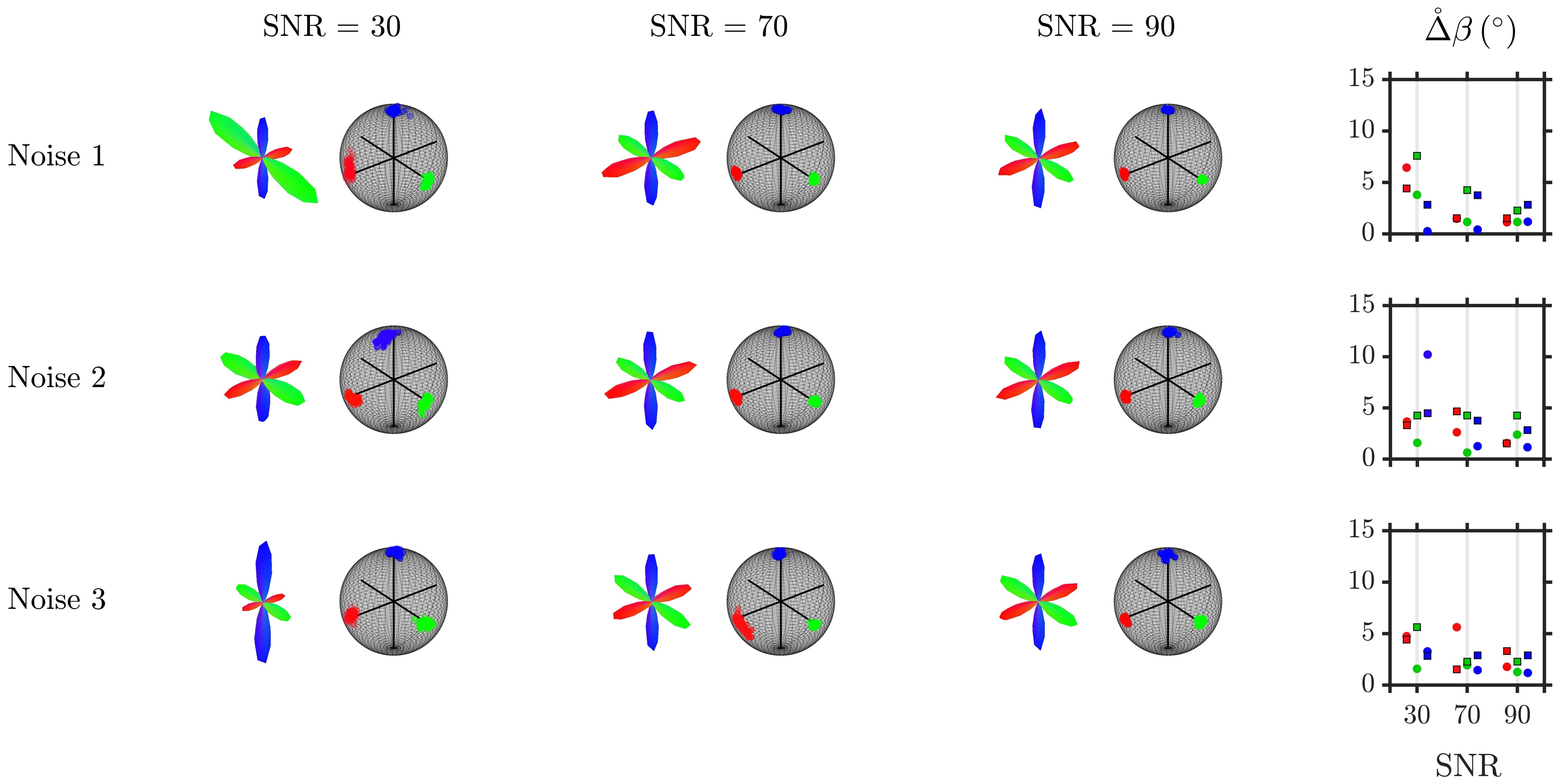}
\caption{Sub-voxel orientations retrieved for three noise realizations of the \textit{in silico} data dubbed "configuration 1" in Section~\ref{Sec_in_silico_data} using the Monte-Carlo inversion for $N_\mathrm{b}=100$ bootstrap solutions and various SNR levels. While the ODFs were obtained via the process detailed in Section~\ref{Sec_ODF_peaks}, the orientational clusters, here represented on the unit sphere, were extracted \textit{via} MC-DPC according to Section~\ref{Sec_MC-DPC}. $\mathring{\Delta}\beta$ denotes the angular deviation, computed for a given orientational cluster as the shortest angle between either the cluster geometric median orientation (circles, see Equation~\ref{Eq_geometric_median}) or the ODF peak (squares, see Section~\ref{Sec_ODF_peaks}), and the closest ground-truth anisotropic component orientation. Color/representation conventions are identical to those of Figures~\ref{Figure_ODF_pedagogic} and \ref{Figure_MC-DPC}. The conditions of the \textit{in vivo} study presented in Section~\ref{Sec_in_vivo_results} are closest to the $\mathrm{SNR} =90$ case.}
\label{Figure_in_silico_orientations_noise}
\end{center}
\end{figure*}

\begin{figure*}[ht!]
\begin{center}
\includegraphics[width=30pc]{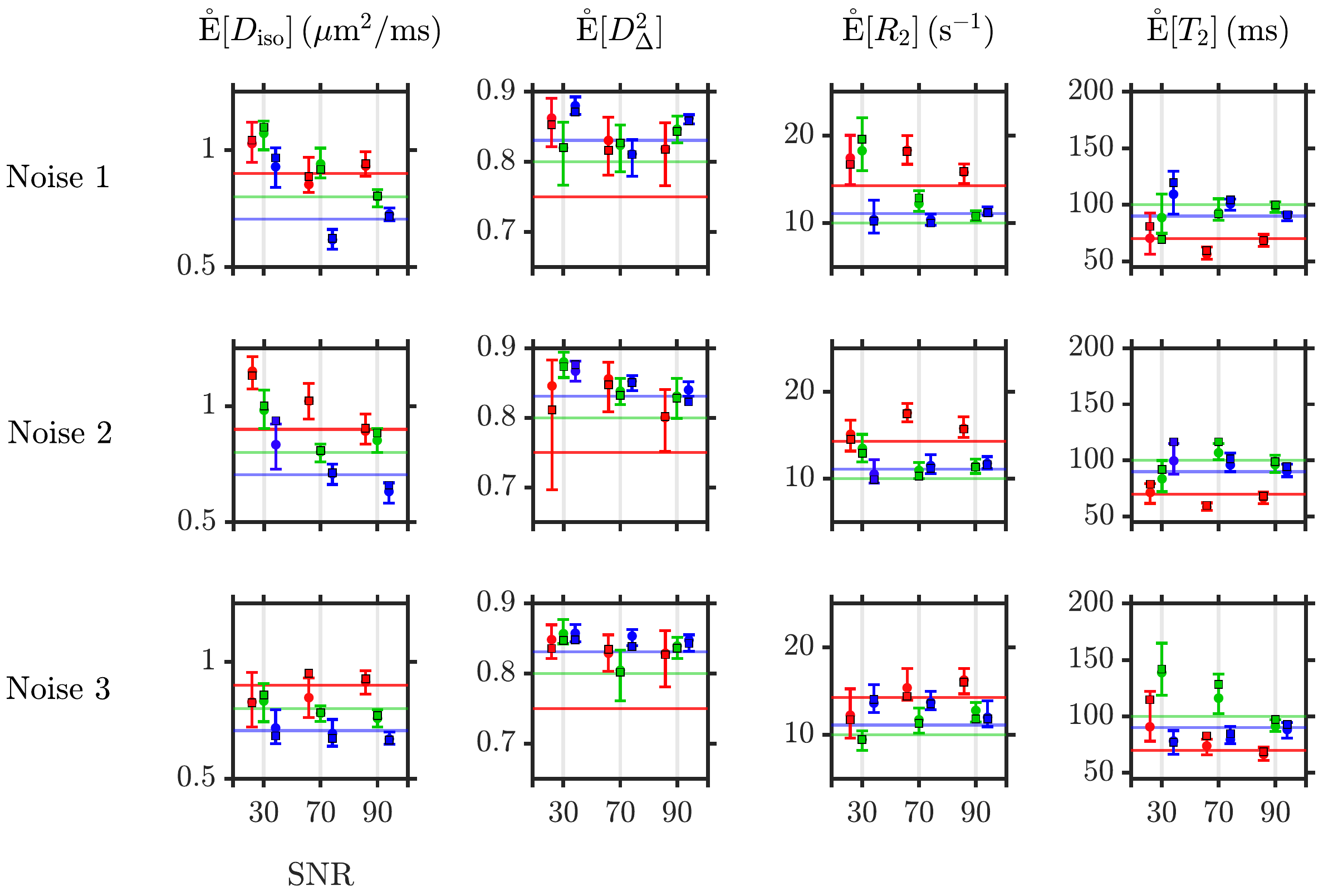}
\caption{Orientation-resolved means $\mathring{\mathrm{E}}[\chi]$ associated with the MC-DPC clusters of Figure~\ref{Figure_in_silico_orientations_noise}. While ground-truth is shown as horizontal lines, the circles and whiskers represent the medians and interquartile ranges of the orientation-resolved means across bootstrap solutions, respectively. Squares correspond to the estimated ODF-peak metrics (see Equation~\ref{Eq_ODF_metrics}, evaluated at an ODF-peak orientation). Colors match those of the orientational clusters/ODF peaks presented in Figure~\ref{Figure_in_silico_orientations_noise}. The conditions of the \textit{in vivo} study presented in Section~\ref{Sec_in_vivo_results} are closest to the $\mathrm{SNR} =90$ case.}
\label{Figure_in_silico_metrics_noise}
\end{center}
\end{figure*}

Figure~\ref{Figure_in_silico_orientations_noise} displays the sub-voxel orientations retrieved for three noise realizations of the \textit{in silico} data dubbed "configuration 1" in Section~\ref{Sec_in_silico_data} using both the ODFs of Section~\ref{Sec_ODF_peaks} and MC-DPC from Section~\ref{Sec_MC-DPC}, with $N_\mathrm{b}=100$ bootstrap solutions. Figure~\ref{Figure_in_silico_metrics_noise} quantifies the orientation-resolved means $\mathring{\mathrm{E}}[\chi]$ for the same sub-voxel orientations. These figures confirm the fact mentioned in Section~\ref{Sec_in_silico_results}, namely that the accuracy and precision of MC-DPC's estimations, reflecting those of the distributions output by the 5D Monte-Carlo signal inversion, are inconsistent at low to intermediate SNR given the acquisition scheme described in Section~\ref{Sec_in_vivo_data}.

\end{appendices}

\vspace*{\fill}

\newpage 
\clearpage
\newpage

\section*{Acknowledgement} 
This work was financially supported by the Swedish Foundation for Strategic Research (ITM17-0267) and the Swedish Research Council (2018-03697). Data collection was approved by the IRB of Cardiff University School of Medicine. D. Topgaard owns shares in Random Walk Imaging AB (Lund, Sweden, \href{http://www.rwi.se/}{http://www.rwi.se/}), holding patents related to the described methods. D. K. Jones and C. M. W. Tax were supported by a Wellcome Trust Investigator Award (096646/Z/11/Z), C. M. W. Tax by a Sir Henry Wellcome Fellowship and a Veni grant (17331) from the Dutch Research Council (NWO), and D. K. Jones by a Wellcome Trust Strategic Award (104943/Z/14/Z).


\end{document}